\documentclass[10pt,journal,compsoc]{IEEEtran}
\IEEEoverridecommandlockouts
\usepackage{booktabs}
\usepackage{amsmath,amssymb,amsfonts}
\usepackage{graphicx}
\usepackage{textcomp}
\usepackage{xcolor}
\usepackage{colortbl}
\usepackage{makecell}
\usepackage{rotating}
\usepackage{multirow}
\usepackage{tabularx}
\usepackage{float} 
\usepackage{subfigure} 
\usepackage{diagbox}
\usepackage{listings}
\usepackage[flushleft]{threeparttable}
\usepackage[justification=centering]{caption}
\usepackage{CJK}
\usepackage{cite}
\usepackage{enumitem}
\setitemize[0]{leftmargin=15pt}
\usepackage[many]{tcolorbox}
\usepackage{xspace}

\usepackage{algorithm}
\usepackage{algpseudocode}
\usepackage{amsmath}

\usepackage[linkcolor=red,anchorcolor=green,citecolor=blue]{hyperref}
\def\BibTeX{{\rm B\kern-.05em{\sc i\kern-.025em b}\kern-.08em
    T\kern-.1667em\lower.7ex\hbox{E}\kern-.125emX}}
    
\usepackage{balance}

\definecolor{graytbl}{gray}{.8}
    
\newcommand\scalemath[2]{\scalebox{#1}{\mbox{\ensuremath{\displaystyle #2}}}}

\newcommand{\tool}{\textit{MLCDroid}\xspace}

\newcommand{\step}[1]{\textcircled{\small{#1}}}
\definecolor{codegreen}{rgb}{0,0.6,0}
\definecolor{codegray}{rgb}{0.5,0.5,0.5}
\definecolor{codepurple}{rgb}{0.58,0,0.82}
\definecolor{backcolour}{rgb}{0.95,0.95,0.92}

\lstdefinestyle{mystyle}{
  frame=single,
  commentstyle=\color{codegreen},
  keywordstyle=\color{magenta},
  numberstyle=\tiny\color{codegray},
  stringstyle=\color{codepurple},
  basicstyle=\ttfamily\scriptsize,
  breakatwhitespace=false,         
  breaklines=true,                 
  captionpos=b,                    
  keepspaces=true, 
  keywordstyle = \color{blue},
  morekeywords = {if}
  numbers=left,                    
  numbersep=6pt,                  
  showspaces=false,                
  showstringspaces=false,
  showtabs=false,                  
  tabsize=2
}

\begin{document}

\title{Multi-label Classification for Android Malware Based on Active Learning}

\author{
Qijing Qiao,
Ruitao Feng,
Sen Chen,~\IEEEmembership{Member,~IEEE},
Fei Zhang,
Xiaohong Li,~\IEEEmembership{Member,~IEEE}
\IEEEcompsocitemizethanks{
\IEEEcompsocthanksitem Q. Qiao and R. Feng (co-first) contributed equally to this paper.
% \IEEEcompsocthanksitem S. Chen is the corresponding author.
\IEEEcompsocthanksitem S. Chen and X. Li are the corresponding authors.
\IEEEcompsocthanksitem Q. Qiao, S. Chen, F. Zhang, X. Li are with the College of Intelligence and Computing, Tianjin University, China.\protect\\
E-mail:\{qqj, senchen, zhangfei, xiaohongli\}@tju.edu.cn
\IEEEcompsocthanksitem R.Feng is with the CSCRC, University of New South Wales, Australia, and the SPIRIT, Nanyang Technological University, Singapore.\protect\\
E-mail:ruitao.feng@unsw.edu.au, feng0082@ntu.edu.sg
}}

\markboth{Journal of IEEE Transactions on Dependable and Secure Computing,~Vol.~xx, No.~xx, xx~2022}%
{Shell \MakeLowercase{\textit{et al.}}: Bare Demo of IEEEtran.cls for Computer Society Journals}

\IEEEtitleabstractindextext{%
\begin{abstract}
The existing malware classification approaches (i.e., binary and family classification) can barely benefit subsequent analysis with their outputs. Even the family classification approaches suffer from lacking a formal naming standard and an incomplete definition of malicious behaviors. More importantly, the existing approaches are powerless for one malware with multiple malicious behaviors, while this is a very common phenomenon for Android malware in the wild. So that both of them actually cannot provide researchers with a direct and comprehensive enough understanding of malware.

In this paper, we propose \tool, an ML-based multi-label classification approach that can directly indicate the existence of pre-defined malicious behaviors. With an in-depth analysis, we summarize 6 basic malicious behaviors from real-world malware with security reports and construct a labeled dataset. We compare the results of 70 algorithm combinations to evaluate the effectiveness (best at 73.3\%). Faced with the challenge of the expensive cost of data annotation, we further propose an active learning approach based on data augmentation, which can improve the overall accuracy to 86.7\% with a data augmentation of 5,000+ high-quality samples from an unlabeled malware dataset. This is the first multi-label Android malware classification approach {intending} to provide more information on fine-grained malicious behaviors.
\end{abstract}

}

\maketitle

\IEEEraisesectionheading{\section{Introduction}\label{intro}}
%background
\IEEEPARstart{C}urrently, Android is the most popular operating system (OS) on mobile devices with over 2.5 billion active users spanning over 190 countries~\cite{Android2021Statistics,chen2019storydroid,chen2022storydistiller}. 
%The market with huge potential brought by Android attracts a large number of developers. 
In 2021, the total number of applications (apps) on Google Play has reached 2.8 million~\cite{Android2021AppBrain}. However, various apps brought not only convenience but also security threats through Android malware. A recent statistic shows the number of new malware amounted to 0.48 million per month~\cite{Android2021statista}. That large number of malware directly or indirectly causes immeasurable harm to users' privacy and property~\cite{44worries2022,tang2019large}. So the classification of Android malware is becoming more and more important.

Existing approaches for Android malware detection include traditional solutions (signature-based method~\cite{zhou2013fast}, behavior-based method~\cite{tam2015copperdroid,meng2018droidecho}, and data flow analysis-based method~\cite{arzt2014flowdroid}) and machine learning (ML)-based solutions~\cite{aafer2013droidapiminer,arp2014drebin,chen2016stormdroid,mariconti2016mamadroid,fan2018android,chen2018automated, feng_mobidroid_2019,fan2016frequent,bai2020icse,feng_seqmobile_2020,feng_performance_sensitive_2021,wu2021android}. Compared with the former one, ML-based method is considered as one of the most promising techniques and achieves high detection accuracy.
ML-based method can be mainly divided into two categories, binary classification~\cite{aafer2013droidapiminer,chen2016stormdroid,arp2014drebin,feng_performance_sensitive_2021,feng_seqmobile_2020,mariconti2016mamadroid,chen2018automated} and family classification~\cite{fan2018android,fan2016frequent,bai2020icse}.
Through the classic binary classification, we can successfully identify whether a certain application is {malicious or benign}. However, even if it is correctly classified, we still have no idea about its specific attack chain and corresponding malicious behaviors. In other words, the detection result can hardly provide any clue for security analysts, who aim at completing the knowledge of the detected 0-day attack efficiently and further {provide} potential victims with alerts and emergency protections.
On the other hand, family classification's focal point is classifying malware into a certain family, whose definition can tell the key malicious behavior the malware may perform and the potential hazards the malware may have.
However, along with the quick iterations on malware, existing family classification methods have two fatal limitations on their engineering basis, which severely undermine their capability towards providing more detailed and useful information for further analysis.
Firstly, by investigating public malware family information{~\cite{maggi2011finding,2010The,10.1007/978-3-319-45719-2_11,2017Euphony,10.1007/978-3-319-40667-1_8}}, we notice there is no common agreement on malware naming or review procedure to oversee the family names. In other words, different agencies may categorize the same malware into different families. It may confuse researchers who attempt to provide a quick response to fight against the malware. Secondly, when analyzing the detailed malicious behaviors in real cases, we have also observed the behaviors in many malware are actually out of their family definitions. {The reason behind this phenomenon is that these are mostly defined according to the original 0-day malware, which may only contain a specific malicious behavior. Hence, when some more variants are developed, even a successful family classification can become useless since most malware family definitions can cover only a specific and limited set of malicious behaviors.}
For those widely existing variants constructed with multiple behaviors, the result of family classification can fail to provide expected information. If we want to understand all the involved malicious behaviors, a great effort in the additional analysis is always necessary.

Considering all the above limitations, we conduct a survey on learning-based classification problems in other domains, such as computer vision, to seek potential inspirations that can enable a stronger capability on providing more detailed malicious information in classification results, especially, for those malware {containing} multiple malicious behaviors. We find our task, which is locating the multiple malicious behaviors in malware as fully as possible, is quite similar to the multi-target detection~\cite{boutell2004learning,kang2006correlated}, which aims to locate all wanted targets in a picture.
Both of them have the same task which is accurately locating multiple uncorrelated targets in a problem space. In the field of multiple malicious behavior detection, the ``uncorrelated targets'' are the malicious behaviors, and the ``problem space'' is the malware.
Therefore, referring to the method used in multi-target detection, we try to adopt a multi-label learning method to provide a more comprehensive classification that has the potential to overcome the above limitations.
In a multi-label classification (MLC) problem, multiple labels are assigned to each classifiable instance~\cite{tsoumakas2007multi}. That means a single malware can be associated with a set of labels that represents different malicious behaviors simultaneously. So, if we can summarize these labels as reasonable and scientific as possible, a direct and full understanding on the malicious behaviors of a certain malware can be promised through the classification results.
In addition to the exciting prospect, applying multi-label classification will also pose new challenges. 
Firstly, as far as we investigated, there is no formal general standard that aims to help in categorizing the detailed malicious program behaviors. That means we have to make relatively precise definitions at the program level for the multi-label classification.
Secondly, despite there are usable Android malware datasets, however, the effort on analyzing and labeling them with precise labels at the program level can be extremely costly. Hence, achieving reasonable usability with a small labeled dataset and discovering an effective way to expand the size of the usable dataset are the other essential problems that need to be handled.

In this paper, we propose \tool, an ML-based multi-label malware classification approach that can point out 6 types of fine-grained malicious behaviors in the classification results. To ensure the effectiveness of the predefined labels, we first summarize 6 types of basic malicious behaviors with the help of an in-depth analysis of a dataset of 180 malware samples with security reports. We then propose 6 inductive labels with a summarized feature dictionary and construct a new labeled dataset. Further, in order to enhance the completeness of the feature dictionary, we supplement it with relevant knowledge from various research papers and technical documents and finally obtain 531 features associated with 6 types of malicious behaviors. To investigate the effectiveness of potential multi-label ML algorithms, we evaluate 70 combinations of multi-label classification and basic classification algorithms on the labeled dataset. To address the challenge of data shortage and further enhance the effectiveness and reliability of our approach, we propose a method called \textit{Detection-Training} by leveraging active learning. According to the evaluation results on diverse ML algorithms, we select 10 trained ML models that yield the best results and use them as the base models. By adopting a data augmentation method using the base models and unlabeled datasets, we not only enlarge the labeled dataset and obtain a better MLC model, but also validate the possibility of active learning in the malware classification domain towards understanding and explaining the inner malicious behaviors.
Through our approach, we finally obtain a relatively high accuracy (i.e., CDN+J48: 86.7\% on the DREBIN dataset and 83.3\% on the VirusShare dataset) in our multi-label classification task. Compared with the accuracy of the base model constructed with the same algorithms (73.3\%), the proposed Detection-Training outperforms with the help of a successful data augmentation of 4,840 additional unlabeled malware samples from the DREBIN dataset and 4,992 from VirusShare dataset.

In summary, we make the following main contributions:
\begin{itemize}
\item {Generally, we propose \tool, an approach that can perform multi-label classification on Android malware which can categorize them into fine-grained malicious behaviors.}
\item {We perform an in-depth study in the field of Android malware to get a well pre-understanding of the malicious behaviors. With that in mind, we summarize 6 types of basic malicious behaviors and define a set of 6 inductive labels by manually analyzing 180 malware with security reports.\footnote{We release the labeled malware as a benchmark (\url{https://github.com/qqj1130247885/MLC-for-Android-Malware}).}}
\item {We basically select the combinations of multi-label algorithm and basic machine learning algorithm which are most suitable for our task by evaluating 70 combinations on the labeled dataset.}
\item {Facing the challenges of the expensive cost on data annotation, we propose an active learning method, called Detection-Training, to enhance the classification capability with data augmentation from the unlabeled dataset. In this way, we can not only improve the effectiveness but also obtain auto-labeled high-quality malware.}
\end{itemize}

\begin{figure*}[t]
\centering
\includegraphics[scale=0.4]{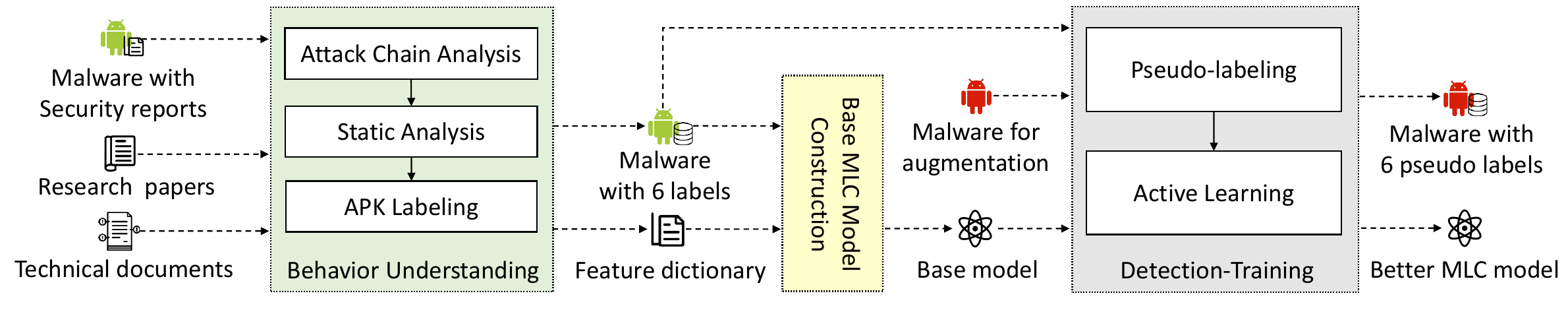}
\caption{An overview of \tool}
\label{fig:overview}
\end{figure*}

\section{Approach}\label{approach}

{\ref{fig:overview} presents the workflow of \tool can be divided into 3 main phases:
(1) Behavior analysis, feature selection, and data annotation: 
We propose 6 different behavior labels as a set with an in-depth analysis on the dataset of malware with security reports.
Next, we summarize the feature dictionary with the help of static analysis and supplement it according to the relevant research papers and technical documents.
After that, we manually label the malware with substantial efforts according to the observed malicious behaviors.
(2) Base MLC model construction:
By comparing the effectiveness of various multi-label classification and basic ML classification algorithms on the labeled dataset, we obtain a base model which performs the best among all candidates.
(3) Detection-Training:
With the obtained base model, we adopt an active learning method based on data augmentation to enlarge the labeled dataset and further improve the model accuracy. 
The details of the above three phases are introduced in the following sections respectively.}

\subsection{Behavior Analysis, Feature Selection, and Data Annotation}\label{approach:behavior_label}

\begin{figure}[t]
\centering
\includegraphics[scale=0.4]{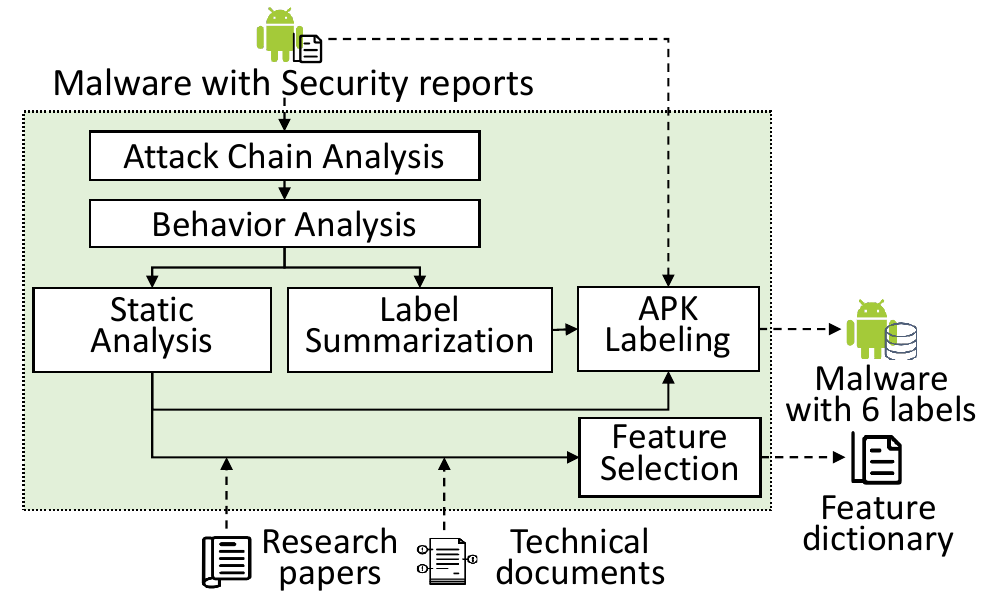}
\caption{Behavior analysis and data annotation}
\label{fig:behavior anlysis}
\end{figure}

%read reports & attack chain analysis
{In order to gain a better understanding and further provide a comprehensive representation of the various malicious behaviors, we first conduct an in-depth analysis of real-world malware with the help of their security reports. {These malware were collected in the period of 2008 to 2018.}
% , nearly half of the samples were collected in 2008.
We collect these data from our industrial partner, anonymousCERT. Note that the security reports have been thoroughly validated and identified by the employees in anonymousCERT. Based on the analysis results, we then summarize 6 types of malicious behaviors. By applying an additional supplementation according to relevant research papers and technical documents, we propose a novel multi-label classification benchmark~\cite{github-BenchMark} with a summarized feature dictionary.
The detailed workflow is shown in Fig.~\ref{fig:behavior anlysis}.}

\subsubsection{Attack chain and behavior analysis}
\label{approach:behavior_label:behavior_analysis}
{To investigate every detail of malicious behaviors for ensuring the coverage of various types, we first study their attack chains by analyzing the corresponding security reports attached to the malware dataset. In this step, we mainly focus on the triggering events, violated assets, and suspicious actions which can help us understand the attack mechanism and actual flow in the implementation. Next, according to this key information, we further categorize the malicious behaviors with their functionalities and associated elements.}

{While performing behavior analysis, we also refer to the family classification results from VirusTotal~\cite{0VirusTotal}. By comparing our behavior analysis with those results, we notice an interesting conflict, which is the malicious behaviors in many malware that are actually out of their family definition.
For example, a malware named ``personal identity electronic certificate'' is classified into the family named ``Android:SMSThief'' by Avast{, AVG, and Alibaba, it is also classified into the family named ``Android.SmsSpy'' by DrWeb and F-secure.}
Generally, malware belonging to ``Android:SMSThief'' may contain the following malicious behaviors, such as ``read mobile phone status'', ``receive, read, write or send text messages'', ``access network connections'', ``access to contact information'', etc.
{Malware belonging to ``Android.SmsSpy'' contains similar malicious behaviors to those belonging to ``Android:SMSThief'', and can also access the WiFi network status information of the mobile phone, etc.}
However, by comparing with our analysis results, we find it also performs other malicious behaviors out of its family definition. In Listing.~\ref{listing:multiple_behaviors}, the highlighted contents show that this malware can monitor the incoming and outgoing phone calls and make phone calls at run-time without the victim's authorization. Besides, it can also set up call forwarding. Obviously, this kind of situation may greatly affect family classification methods on their effectiveness, and even seriously mislead further analysis and behavioral interpretation.
Therefore, a new classification standard that can provide an essential explanation on more precise knowledge of basic malicious behaviors is much needed.}

\begin{lstlisting} [float,floatplacement=H,numbers=left, basicstyle=\scriptsize,caption={{A sample: multiple behaviors in one malware}}, label={listing:multiple_behaviors}]
<@\textcolor{lightgray}{//Relevant permissions extracted from AndroidManifest.XML}@>
<@\textcolor{codepurple}{android.permission.READ\_PHONE\_STATE}@>
<@\textcolor{red}{android.permission.RECEIVE\_SMS}@>
<@\textcolor{red}{android.permission.READ\_SMS}@>
<@\textcolor{red}{android.permission.WRITE\_SMS}@>
<@\textcolor{red}{android.permission.SEND\_SMS}@>
<@\textcolor{blue}{android.permission.CALL\_PHONE}@>
<@\textcolor{brown}{android.permission.INTERNET}@>
<@\textcolor{teal}{android.permission.READ\_CONTACTS}@>

<@\textcolor{lightgray}{//Relevant implementation extracted from source code}@>
<@\textcolor{lightgray}{//Behavior of retrieving user's SMS}@>
private void getSMS(){
    Cruse v10=this.<@\textcolor{red}{getContentResolver().qury}@>(Uri.parse(``Content://sms/inbox''),v2,(String)v2),v2,(String)v2));
    ...
}
<@\textcolor{lightgray}{//Behavior of uploading user's SMS to a remote server}@>
private void run(){
    ...
    Log.i(MainActivity.this.TAG,MainActivity.this.<@\textcolor{brown}{httpPOST}@>(CommonMoudle.Url7,((List)v0)));
    ...
}
<@\textcolor{lightgray}{//Behavior of monitoring user's outgoing calls}@>
public void onReceive(Context arg7,Intent arg8){
    ...
    if("<@\textcolor{teal}{Android.intent.action.NewOutgoingCall}@>".equal(arg8.getAction())){
        ...
    } 
    ...
}
<@\textcolor{lightgray}{//Behavior of making phone call}@>
private void quickCall(){
    ...
    v2.setAction("<@\textcolor{blue}{android.intent.action.CALL}@>");
    ...
}
<@\textcolor{lightgray}{//Behavior of setting up call forwarding}@>
private void transferredCall(){
    ...
    v4.invoke(this.telephonyManager,null).<@\textcolor{codepurple}{endCall()}@>;
    ...
    v2.setAction("<@\textcolor{codepurple}{Android.intent.action.Call}@>");
    ...
}
\end{lstlisting}

\subsubsection{Label definition}
\label{approach:behavior_label:label_definition}
{With the in-depth behavior analysis on the collected malicious apps and their security reports, we summarize 6 types of basic fine-grained malicious behaviors and define 6 inductive labels, which are ``\textbf{SMS-related}'', ``\textbf{Internet-related}'', ``\textbf{Telephony-related}'', ``\textbf{Lock-in}'', ``\textbf{Ads}'', and ``\textbf{Re-infection}''}. A brief description of each label is provided in Table~\ref{tab: sample number of each class}. {These 6 types are the most representative and can cover all the malicious behaviors that occur in our dataset.} The detailed introduction of their mechanisms and potential threats are shown as follows:

\begin{itemize}
\item{\textbf{SMS-related:}
{This type of behavior can cause various cyber-attacks conducted via SMS service, such as privacy leakage and service charge.
The first case type can silently steal the victim's privacy, such as personal information, digital identities to properties, and contents of the user's private conversations, by sending them from the infected device to a remote server via SMS service.
The second case type can cause unaware service charges by bypassing a two-factor authentication (2FA), such as silently confirming a subscription payment by reading the verification code in messages at runtime.
The third case type can perform SMS-based malicious behaviors with remote control commands received from a remote server via SMS service. 
The fourth case type can delete the messages in the user's SMS box.}}

\item{\textbf{Internet-related:}
{This type of behavior can send the user's privacy from a local device to the attacker's remote server by requesting network connections.
In our identified cases, the attack actions are performed according to the control instructions received from the server. 
}}

\item{\textbf{Telephony-related:}
{This type of behavior can perform telephone call-related attacks which can cause privacy leakage or economic losses.
The first case type can steal the victim's telephone call-related information, such as incoming \& outgoing call history and saved contact information.
The second case type can monitor the incoming \& outgoing calls at runtime. A highly representative case is that some ransomware only allows victims to answer the designated call on the infected device.
}}

\begin{table}[t]
\caption{An overview of ground-truth dataset}
\label{tab: sample number of each class}
\centering
\begin{tabular}{|p{0.09\textwidth}p{0.24\textwidth}c|p{0.04\textwidth}}
\hline
\textbf{Label }              & \textbf{Description }                                                                                                                                            & \textbf{\# of samples} \\ \hline\hline
SMS-related         & This type of behavior can cause various cyber attacks conducted via SMS service, such as privacy leakage and service charge.                                                                                  & 113           \\ \hline
Internet-related    &   This type of behavior can send the user's privacy from a local device to the attacker's remote server by requesting network connections.     & 50            \\ \hline
Telephony-related  & This type of behavior can perform telephone call-related attacks which can cause privacy leakage or economic losses.                                                          & 10            \\ \hline
Lock-in  & This type of behavior can extort money by blocking the victim's control behaviors.                        & 73            \\ \hline
Ads                 & This type of behavior usually distributes annoying advertisements, such as pornographic and gambling websites, to victims and possibly causes economic losses. & 13            \\ \hline
Re-infection & This type of behavior can prevent itself from being uninstalled so that it can keep performing attack actions on the infected device.        & 67            \\ \hline
\end{tabular}
% }
\end{table}

\item{\textbf{Lock-in:}
{This type of behavior can extort money by blocking the victim's control behaviors.
The first case type can bind its own layout on the top layer of the UI (a.k.a., user interface) after installation unless the victim pays the ransom according to their instructions. 
The second case type can modify the lock screen password without the user's authorization. Same as the first case, it also asks the user to transfer money to a designated e-money wallet. Once the payment is confirmed, the unlock password will be sent to the victim.
Besides, some variants in this type can change the password and lock the phone repeatedly, even though the user has paid the ransom many times.}}

\item{\textbf{Ads:}
{This type of behavior usually distributes annoying advertisements, such as pornographic and gambling websites, to victims and possibly causes economic losses.
The first case type displays advertisements at the top layer of the UI by frequently generating pop-up windows. Once the victim misclicks them, it will automatically switch to another app or website.
The second case type can send the download URL to all saved contacts via group SMS service.}}

\item{\textbf{Re-infection:}
{This type of behavior can prevent itself from being uninstalled so that it can keep performing attack actions on the infected device.
In our investigated cases, it always exists in conjunction with other types of malicious behaviors. When installed by the victims, the malware can induce them to activate the device management authority and hide the desktop icon, so that the program cannot be uninstalled normally. At the same time, some variants will also monitor operations like mobile phone startup, screen brightening, text messages, phone calls, etc., and keep running in the background after self-starting, so as to achieve the purpose of persistence.}}
\end{itemize}

\subsubsection{Feature selection}
\label{approach:behavior_label:dict_construction}
{{In order to extract the features which can represent malicious behaviors in malware properly, we build a feature dictionary. We select 3 types of widely-used features~\cite{feng_performance_sensitive_2021,Ali2017AndroDialysis}, which are API, used permission, and intent.}
API is most important in our approach since it is the core component in malicious behaviors' demonstration and a procedure call interface to operating system resources. Used permission is the relevant access right that needs to be applied before performing malicious behaviors. Intent provides a mechanism to assist in the interaction and communication between activities, acting as an intermediary.}

{With the knowledge of the 6 defined types of malicious behaviors, we first summarize the key program-level components of each type from the security reports.
Since the analysis-oriented and discriminating basis of relevant features is highly associated with the knowledge of malicious behaviors and defined behavior labels, it is essential to validate the concrete implementation step by step, in case any mistake or missing information which may cause fatal damage to the usability and effectiveness of our approach.
Thus, we further perform a validation based on static analysis.
After decompiling the samples back into source code, we first locate the potential candidates of the malicious behaviors by searching the key APIs or the statement of permission requests and intent actions. To further ensure the completeness and correctness of our summarized key features, we then validate the candidates by analyzing the program flow of control at the level of API call.} {After this process, the size of the feature dictionary is 401, including 109 APIs, 59 used permissions, and 23 intents.}
%\sen{How many features in current status?}

\begin{lstlisting} [float,floatplacement=H,numbers=left, basicstyle=\scriptsize, caption={{A sample: an SMS-related malicious behavior}}, label={listing:SMS_behavior}]
<@\textcolor{lightgray}{//Relevant permissions extracted from AndroidManifest.XML}@>
<@\textcolor{blue}{android.permission.READ\_SMS}@>
<@\textcolor{blue}{android.permission.WRITE\_SMS}@>
<@\textcolor{blue}{android.permission.SEND\_SMS}@>

<@\textcolor{lightgray}{//Relevant implementation extracted from source code}@>
public void onReceive(Context arg0, Intent arg11)
{   
    ...
    else if(arg11.getAction().equals("<@\textcolor{brown}{android.intent.action.SMS\_RECEIVED}@>"){
        ...
        StringBuffer v1=new StringBuffer(v8[v7].<@\textcolor{red}{getOriginatingAddress()}@>);
        if(v1.toString().contains("10658166")){
            this.<@\textcolor{red}{abortBroadcast()}@>;
            SmsManager.getDefault().<@\textcolor{red}{sendTextMessage}@>(v1.toString(),v2,"Y",(PendingIntent)v2),(PendingIntent)v2)); 
        ... 
        }
    }
}
\end{lstlisting}

{Listing.~\ref{listing:SMS_behavior} demonstrates a malicious code snippet about the malicious behavior that sending confirmation SMS without authorization in the malware named ``passion movie''. It first obtains the permissions related to SMS, and then keeps monitoring the SMS inbox by continuously checking the actions performed by the victim's device. If the current captured action equals to \textit{android.intent.action.SMS\_RECEIVED}, the malware is aware that there is an SMS arrived at the inbox.
Once received any message, it captures sender's number by calling the function \textit{getOriginatingAddress()}. If the number equals to the designated number ``10658166'', it will call the function \textit{abortBroadcast()} to block the incoming SMS notification, and then replies a ``Y'' as a confirmation message silently. Through these above actions, the service subscription will be confirmed without the user's knowledge. By analyzing the behaviors along the malware's attack chain, we can successfully summarize and supplement the relevant permissions, intents, and APIs to our feature dictionary.}

{While analyzing detailed cases, we find there exist diverse implementations for a same type of malicious behaviors, which result in different key features. For example, the malicious behavior of sending SMS may be achieved by calling \textit{SendSMS()}, \textit{sendTextMessage()}, \textit{sendMultipartTextMessage()}, or others. However, the limited number of malware with reports seriously weakens the completeness of our summary on the characteristics of malicious behaviors. To handle this problem, we supplement our feature dictionary by referring to some relevant research papers and technical documents~\cite{0docs,bankAPP,feng_performance_sensitive_2021,bai2020icse,fan2018android}. 
For research papers, we locate the relevant behaviors corresponding to the 6 defined labels, and supplement those missing features to our dictionary. For technical documents, we check the descriptions of relevant functionalities to seek potential similar implementation. Once we find any, we will then analyze the relevant features, and add them to the feature dictionary as a supplementation.
{During our study, we find some API differences caused by different Android versions and make some updates to the APIs in the feature dictionary. By complementing the feature set with research papers and technical documents, the size of the feature dictionary increases by 130, among which, APIs increase by 80, permission increases by 20, and intent increases by 30.} Finally, we obtain a relatively comprehensive feature dictionary of 531 features, which includes 189 APIs, 79 used permissions, and 263 intents (more details~\cite{googleSites}).}

\subsubsection{Data annotation}
{With the best understanding of the malicious behaviors, we then label the malware samples according to the analysis results. Each sample has a six-dimensional vector representing the 6 labels. Each dimension of the vector corresponds to a type of malicious behavior which is defined in \S~\ref{approach:behavior_label:label_definition}. If the corresponding behavior type to the dimension exists in the malware sample, the dimension will be marked as ``1'', otherwise, it will remain at ``0''. Table.~\ref{tab: sample number of each class} shows the number of malware samples in each label category. In this work, this well-labeled dataset will serve as the ground truth afterward.}

\subsection{Base MLC Model Construction}\label{approach:base_model}
{{
\begin{figure}[t]
\centering
\includegraphics[scale=0.4]{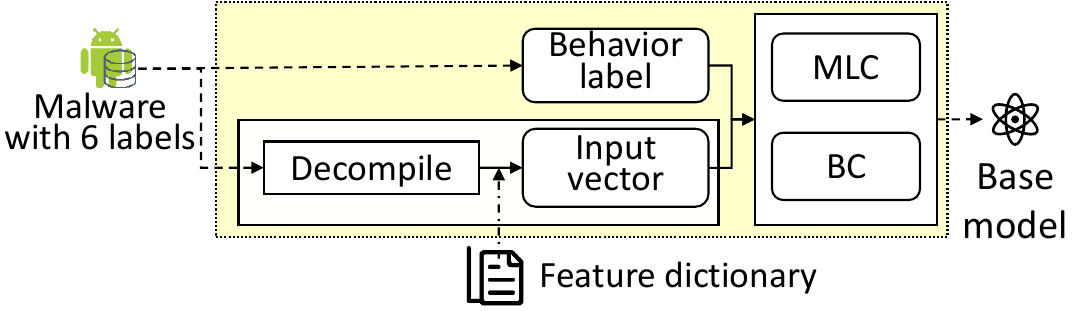}
\caption{Base MLC model construction with labeled dataset}
\label{fig:base model construction}
\end{figure}
}}
{The most well-known two categories of methods in MLC are algorithm adaptation (AA) and problem transformation (PT)~\cite{tsoumakas2007multi, 2020MLC}. AA method uses a variety of algorithms to convert a single-label learning model into a multi-label learning model. PT methods transform a multi-label learning problem into multiple single-label learning tasks.}
{In order to obtain the best base practice, we totally adopt 70 different combinations of MLC and basic classification algorithms to construct ML-based MLC models.}
% mlc algo
{In our work, we adopt 10 different multi-label classification algorithms in total. Table~\ref{tab:MLC_algo} shows the acronym of each algorithm and the type of its belonging method. Totally, there are one algorithm {belonging} to the AA method and 9 algorithms {belonging} to the PT method. 
To the best of our knowledge, these MLC algorithms have never been adopted on any task in Android malware classification.
Besides, we pick 7 different kinds of basic classification algorithms, which are J48, LMT, RandomForest, OneR, PART, RandomTree, and REPTree, and combine them with those MLC algorithms individually.}
\begin{table}
\centering
\caption{MLC algorithms compared in our work}
\label{tab:MLC_algo}
\setlength{\tabcolsep}{1.1mm}{
\begin{tabular}{|lcc|}
\hline
\textbf{Algorithm Name} & \textbf{Acronym} & \textbf{Type}\\ \hline\hline
Binary Relevance & BR & PT \\ \hline
Classifier Chain & CC & PT \\ \hline
Random k-Label Disjoint Pruned Sets & RAkELd & PT\\ \hline
Pruned Sets & PS & PT\\ \hline
Pruned Sets with Threshold & PSt & PT \\ \hline
Multi-Label Back Propagation Neural Network & ML-BPNN & AA \\ \hline
Ranking and Threshold & RT & PT   \\ \hline
Conditional Dependency Networks & CDN & PT  \\ \hline
Conditional Dependency Trellis & CDT & PT  \\ \hline
Classifier Trellis & CT & PT \\ \hline
\end{tabular}
}
\end{table}
\begin{figure}[]
\centering
\includegraphics[width=\linewidth]{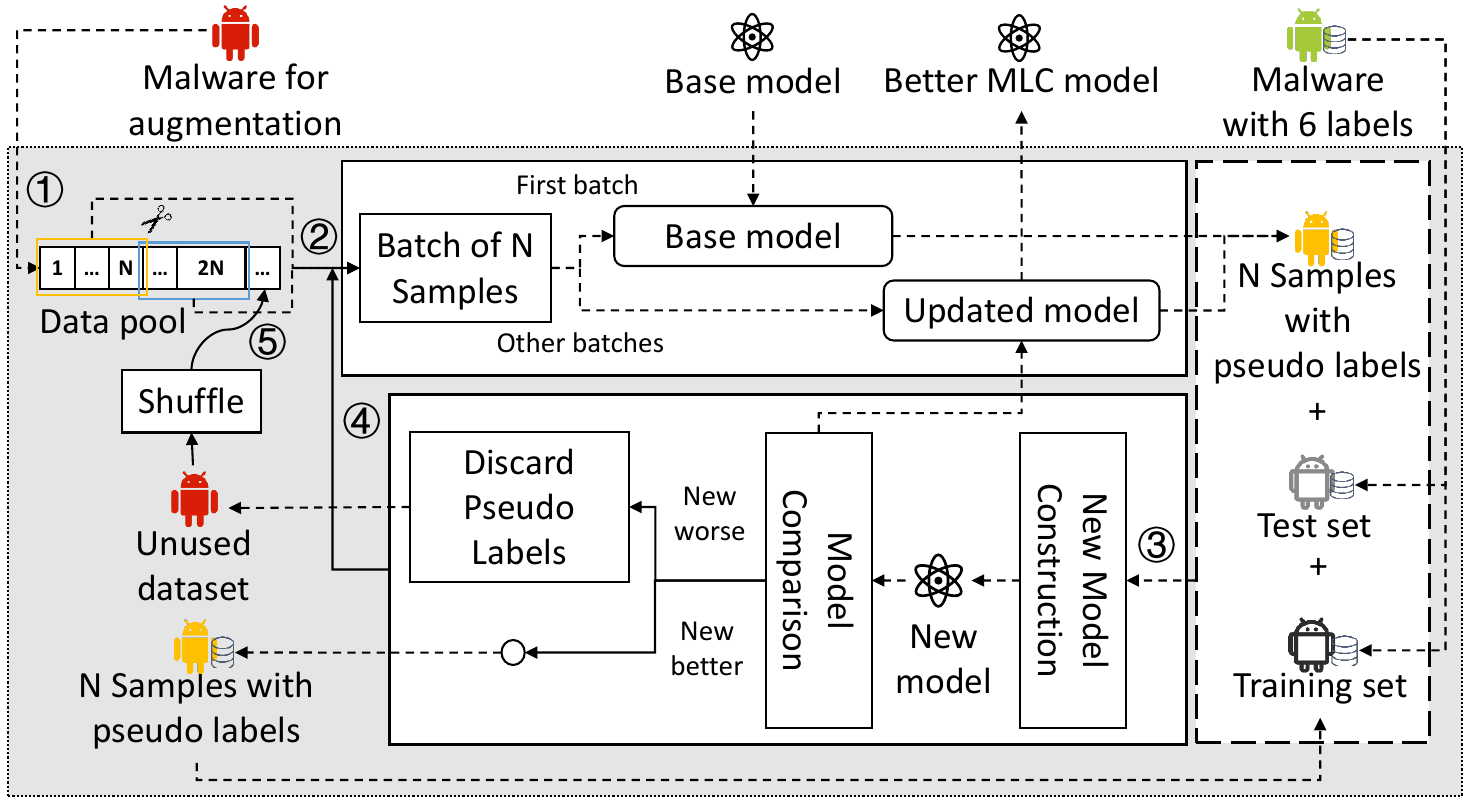}
\caption{{Detection-Training: an active learning framework based on data augmentation}}
\label{fig:active learning}
\end{figure}

% Workflow
{The brief workflow is shown in Fig.~\ref{fig:base model construction}. We first extract feature vectors by decompiling and analyzing the apps in our labeled dataset, according to the feature dictionary constructed in the previous phase. Together with the predefined labels, the input vectors are then fed into each pre-implemented combination of MLC and basic classification (a.k.a, BC) algorithms to train the candidates for the base model.}

\subsection{Detection-Training}\label{approach:detection_training}
{For applications in the data science field, a large-scale high-quality dataset is always the most fundamental basis to ensure completing the established tasks at a promising quality. However, collecting and organizing a new labeled dataset is time-consuming and labor-intensive. In the cybersecurity domain, it can be even harder and more expensive. For the solutions in other domains, such as computer vision, when we are dealing with the task of labeling pictures, we can distribute it to anyone non-professional but willing to participate and then validate the result with a simple random sampling. Back to our task of labeling malware, the first challenge is that very fewer people have the qualification to finish the task at a reasonable confidence level. Secondly, even if we have the support of well-trained security analysts, collecting and labeling thousands of malware samples according to their behaviors {may take too
much time}. Hence, despite of there exist some public malware datasets, expanding the size of the labeled dataset is still a critical challenge that we can barely withstand its cost. So that we design and propose an alternative solution, namely \textit{Detection-Training}, inspired by the research work of both active learning~\cite{tran2019bayesian} and data augmentations~\cite{lee2013pseudo,tarvainen2017mean} in other domains.}

{
Fig~\ref{fig:active learning} shows the workflow of Detection-Training, which is an active learning framework based on data augmentation. 
The framework takes the labeled dataset from the first phase described in \S~\ref{approach:behavior_label}, the base model from the second phase described in \S~\ref{approach:base_model}, and an unlabeled malware dataset as inputs. The feature extraction step with our constructed dictionary is considered as default and omitted in Fig.~\ref{fig:active learning}. The output is an MLC model which outperforms the base model.
The detailed procedures are illustrated as follows:}

{\begin{itemize}[leftmargin=*]
    \item In step~\step{1}, the malware dataset without our predefined labels, namely ``Malware for data augmentation'' is divided into several batches in size $N$.
    \item In the next step~\step{2}, the current first batch is picked and removed from the data pool. Then, the current best MLC model, which is the base model for the first batch of data and updated model for other batches, takes the picked batch as input and generates a classification result for each malware sample in this batch. In this way, these results are temporarily determined as basic facts and further paired with the $N$ samples as pseudo labels.
    \item Next, we combine the batch of data with the current training set to train a new model in step~\step{3}. The test set is remaining unchanged for the purpose of evaluating the quality of this new model by comparing it to the current best model.
    When determining the better model, we mainly compare the classification accuracy of the model. In the first epoch, we update the current model only if the new model has a higher accuracy for improving the accuracy as much as possible. 
    In the next epochs, the models with the same accuracy as the current model are also considered in order to obtain a large data augmentation.
    The specific idea behind adopting different methods for the first epoch and others is that improving the capability of classification is always the first-order target when there is any potential for accuracy improvement.
    Besides, an MLC model with higher accuracy can also classify the batches with higher confidence. 
    Through the method, the better new model is used as the updated model to label the next batch of $N$ samples. The associated malware samples are added to the training set permanently. Otherwise, the new model and pseudo labels will be discarded. {The discarded batch of $N$} samples will be marked as unused data.
    \item In step~\step{4}, the workflow loops along step~\step{2} and step~\step{3} until all the malware samples in the data pool have been picked. Because the classification capability of the updated model is increased after many cycles, the updated model will also have a higher possibility to conduct more precise classifications on those unused data and further enhance the detection capability as well.
    \item Besides, in order to avoid the potential influence from the fixed order in the data pool, the unused dataset is shuffled before returning to the end of the data pool in step~\step{5} and fed to the updated model in step~\step{2} to start next epoch until all the samples are labeled or the accuracy and scale of data augmentation are no longer improved.
\end{itemize}}

{Through all the procedures in this framework, both the classification capability of our MLC model and the size of the usable dataset can be significantly improved.}

\section{Evaluation}
\label{evaluation}
{In this section, we first introduce the used datasets, the experiment environment, and the metrics selected for evaluating the classification capability of the MLC model. After that, We evaluate the effectiveness of different MLC and BC algorithm combinations. Later, we investigate the effectiveness of Detection-Training. At last, we perform an experiment for validating the authenticity of the pseudos labels collected through Detection-Training.}

\subsection{Used Datasets}
\label{evaluation:dataset}
\subsubsection{Manually labeled malware}
{This labeled dataset is constructed on 180 malware samples with manual analysis reports, which are obtained from an anonymousCERT. With our effort, each sample is labeled according to the summarized 6 types of malicious behaviors. A more detailed description is shown in Table~\ref{tab: sample number of each class}. We release these labeled data as a benchmark~\cite{github-BenchMark}.}

\subsubsection{DREBIN dataset}
\label{evaluation:dataset:drebin}
{We choose the classic Android malware dataset DREBIN~\cite{arp2014drebin} as the first unlabeled dataset for our experiment of data augmentation. This dataset was collected in the period of August 2010 to October 2012, and it contains 5,560 applications from 179 different malware families. However, during decompiling, these malware and extracting feature vectors, some malware samples failed in prepossessing. The number of usable samples in our experiments is 5,456 in total.}

\subsubsection{VirusShare dataset}
\label{evaluation:dataset:virusshare}
{We also collected {5,500} Android malware samples from VirusShare~\cite{0VirusShare} as the second unlabeled dataset. These samples have been collected in the period of 2018 to 2022.}
{The process of building the VirusShare dataset is shown in \S~\ref{data:virusshare}.}

\subsection{Experimental Environment}
{All experiments are conducted on the Ubuntu 18.04 server with 36 Intel Xeon E5-2699 v3 CPUs and 192GB RAM.}

{The based language of most implementations in our work is Python 3.6.
MEKA v1.9.2~\cite{MEKA,MEKA_1.9.2}, which is designed to provide a series of algorithms and evaluation indicators to solve multi-label classification problems, is chosen as the basic toolkit in training and evaluating the models. It is an open-source project based on the WEKA~\cite{2009The}.}
 
\subsection{Evaluation Metrics}
\label{evaluation:metrics}
{Generally, the evaluation of the multi-label classification approach is much more complicated than binary or family classification~\cite{0multi-metrics}, since there are multiple classes associated with each sample. Besides, for the active learning method, the effectiveness of the method is highly dependent on the correctness of feedback (i.e., generated pseudo labels). Therefore, we adopt 2 categories of metrics for different purposes, which are 4 sample-based metrics for the overall effectiveness of the trained model and a label-based metric for the effectiveness on different labels~\cite{wu2017unified}. Overall, we take the sample-ACC as the first-order metric in all experiments, and the rest are used as auxiliary ones for determining the capability of models towards various factors.}

{To help in understanding the mathematical definitions of the metrics, we first introduce the definitions of all used notations with their values or denotations if exist as shown in Table~\ref{tab:notation_definition}. In order to make it easier to be understood, we explain each notation in more detail and its interrelationship with other notations in the following paragraph.}

{$L$ denotes the 6 predefined labels, and $|L|$ represents the total number of labels (i.e., 6).
$D$ is the test data set used to evaluate the pre-trained models, and $|D|$ represents its total number of samples. $x_i$ represents the $i$th sample in $D$.
$Y_i$ is an $L$-dimension boolean vector $(Y_{i,1}, \cdots, Y_{i,j}, \cdots, Y_{i,|L|})$ which represents the $|L|$ labels of sample $x_i$ in vector space. $Y_{i,j}$ could be either $1$ or $0$, which means the sample is actually relevant or irrelevant to the malicious behavior associated with label $j$.
$Z_i$ is also an L-dimension boolean vector $(Z_{i,1}, \cdots, Z_{i,j}, \cdots, Z_{i,|L|})$ which represents the classification result of sample $x_i$. $Z_{i,j}$ could be either $1$ or $0$, which means the sample is classified into label $j$ by the pre-trained model, or not.
Besides, in Table~\ref{tab:conf_matrix}, we also present the basic logic conditions while calculating confusion matrix for the sample $x_i$ on label $j$, according to the actual label $Y_{i,j}$ and classification result $Z_{i,j}$.}

\begin{table}[t]
\centering
\caption{A summary of basic notations}
\label{tab:notation_definition}
\begin{tabular}{|p{0.05\textwidth}p{0.24\textwidth}c|p{0.22\textwidth}}
\hline
\textbf{Notation}  &   \textbf{Definition}  &   \textbf{Value/Denotation}\\\hline\hline
$L$ & The predefined labels    &   $|L|=6$\\\hline
$D$ & The set of test data   &   N/A\\\hline
$x_i$ & The $i$th sample in test dataset   &   $x_i \in D$\\\hline
$Y_{i,j}$   & The $j$th value in the actual label vector of sample $x_i$   &   $Y_{i,j} \in \{0,1\}$\\\hline
$Y_i$   & The actual label vector of sample $x_i$   &   $(Y_{i,1}, \cdots, Y_{i,|L|})$\\\hline
$Z_{i,j}$   & The $j$th result in the classification result vector of sample $x_i$   &   $Z_{i,j} \in \{0,1\}$\\\hline
$Z_i$   & The classification result vector of sample $x_i$  &   $(Z_{i,1}, \cdots, Z_{i,|L|})$\\\hline
\end{tabular}
\end{table}
\begin{table}[t]
\centering
\caption{The conditions for calculating confusion matrix}
\label{tab:conf_matrix}
\begin{tabular}{|l|l|l|}
\hline
\diagbox{\textbf{Label}}{\textbf{Pred.}}  & \textbf{Pos.}  & \textbf{Neg.}            \\\hline
\textbf{Pos.}  & $(Y_{i,j}=1) \land (Z_{i,j}=1)$ & $(Y_{i,j}=1) \land (Z_{i,j}=0)$           \\\hline
\textbf{Neg.} & $(Y_{i,j}=0) \land (Z_{i,j}=1)$ & $(Y_{i,j}=0) \land (Z_{i,j}=0)$     \\ \hline
\end{tabular}
\end{table}
\subsubsection{Sample-based metrics}
{In all, we adopt 4 sample-based evaluation metrics, they are hamming loss, zero-one loss, F1-score, and sample-ACC}.

\noindent {\textbf{Hamming loss} represents the fraction of the misclassified labels to the total label number among all samples. The smaller the value of hamming loss is, the stronger the classification ability the MLC model has. It can be calculated with the following equation:}

{
$$\scalemath{1}{
\operatorname{hammingLoss} = \frac{1}{|D|\cdot|L|} \sum_{i=1}^{|D|} \sum_{j=1}^{|L|} Y_{i,j} \oplus Z_{i,j};
}$$}

\noindent {Where $\oplus$ stands for the XOR operation in Boolean logic.}

\noindent {\textbf{Zero-one loss} is a metric that evaluates the fraction of misclassified samples to the entire dataset. For a given input $x_i$, this metric considers the classifier makes a correct classification, only if all labels in the classification results $Z_i$ equal to the ground truth $Y_i$. Otherwise, if any single label is different, the classification is considered as a failure. 
The smaller the value of zero-one loss, the better the performance. 
It can be calculated with the following equations:}

{\begin{equation*}
\scalemath{1}{
\left\{
\begin{alignedat}{3}
\begin{split}
& loss_{i}=\left\{\begin{matrix}
1, Y_{i} \neq Z_{i},\\
0, Y_{i} = Z_{i},
\end{matrix}\right.\\
& \operatorname{zero-one\ loss}=\frac{1}{|D|} \sum_{i=1}^{|D|} loss_i;
\end{split}
\end{alignedat}
\right.}
\end{equation*}}

\noindent {Where $loss_i$ denotes the classification result of the sample $x_i$. It could be either $1$ or $0$, which means the sample is correctly predicted or incorrectly predicted.}

\noindent {\textbf{F1-score} measures the accuracy and completeness of the classification. The F1-score adopted in evaluating the proposed multi-label classification approach is calculated by the F-measure averaging on each sample. A higher F1-score means better performance in classification. It can be calculated with the following equation:}

{
$$\scalemath{1}{
\text {F1-score}=\frac{1}{|D|} \sum_{i=0}^{|D|} \frac{2\sum_{j=0}^{|L|} Y_{i,j} \land Z_{i,j}}{\sum_{j=0}^{|L|}Y_{i,j}+\sum_{j=0}^{|L|}Z_{i,j}};
}$$}

\noindent {\textbf{Sample-ACC} is adopted to evaluate the overall accuracy of the pre-trained model on the test dataset. The idea that we are trying to explore is providing as much explanation as possible in order to provide some usable and correct clues for future analysis.
However, different from the method used in binary or family classification tasks, ensuring to correctly classify all $|L|$ labels of the given sample could be extremely challenging in an MLC task. Especially, the limited size of the labeled dataset will definitely make this target become a ``mission impossible''.
Hence, we use an alternate definition of the correctness of the classification result.
We consider the sample $x_i$ to be correctly classified, as long as its classification result satisfies the following two conditions:
\begin{itemize}
    \item Detect at least one malicious behavior (true positive: $(Y_{i,j}=1) \land (Z_{i,j}=1)$) correctly.
    \item Predict nonexistent malicious behaviors (false positive: $(Y_{i,j}=0) \land (Z_{i,j}=1)$) to none.
\end{itemize}

Otherwise, the sample is misclassified.
{In all, the goal is to predict the malicious behavior of the sample as much as possible, meanwhile, avoiding misleading with incorrect information.}
The mathematical definition of the sample-ACC is shown as follows:}

{\begin{equation*}
\scalemath{1}{
\left\{
\begin{alignedat}{3}
\begin{split}
& C_i= \{Z_{i,j} | \urcorner ((Y_{i,j}=0) \land (Z_{i,j}=1)) ,j \in N \cap (0,|L|]\},\\
& C = \{x_i | \forall x_i \in D, C_i \ni 1, i \in N \cap (0,|D|]\},\\
& \operatorname{sample-ACC}=\frac{|C|}{|D|};
\end{split}
\end{alignedat}
\right.}
\end{equation*}}

\noindent {Where $C_i$ denotes the set of classification results of the sample $x_i$ after removing false positives; and $C$ denotes the set of correctly classified samples, while each sample contains at least one successfully detected malicious behavior. Sample-ACC represents the proportion of correct samples in the entire test dataset.}

\subsubsection{Label-based metric}
{To evaluate the detection ability of the MLC model on each label, we also select a label-based metric, namely label-ACC.}

\noindent {\textbf{Label-ACC} aims to evaluate the accuracy of the pre-trained model on the selected labels. In other words, the larger the label-ACC, the better the model performs on the label.
The idea behind this metric is that, because of the limited and imbalanced labeled data, the overall accuracy calculated on all samples may hide the poor effectiveness of the classifier on those labels that contain fewer samples. The mathematical definition of the label-ACC is shown as follows:}

{\begin{equation*}
\scalemath{1}{
\left\{
\begin{alignedat}{3}
\begin{split}
& C_j= \{x_i | \forall x_i \in D, i \in N \cap (0,|D|],\\
&\hspace{35pt}((Y_{i,j}=1) \land (Z_{i,j}=1)) \lor ((Y_{i,j}=0) \land (Z_{i,j}=0))\},\\
% & C_j= \{x_i | ((Y_{i,j}=1) \land (Z_{i,j}=1)) \lor ((Y_{i,j}=0) \land (Z_{i,j}=0)), i \in N, i \in (0,|D|]\},\\
% & D_j= \{x_i | Y_{i,j}=1, i \in N, i \in (0,|D|]\},\\
& D_j= \{x_i | \forall x_i \in D, Y_{i,j}=1, i \in N \cap (0,|D|]\},\\
& \operatorname{label-ACC}=\frac{|C_j|}{|D_j|};
\end{split}
\end{alignedat}
\right.}
\end{equation*}}

\noindent {Where $C_j$ denotes the set of correctly classified samples on label $j$. Here, the basis of judgment includes both successfully detected samples containing malicious behavior and successfully detected samples that do not contain malicious behavior. $D_j$ denotes the set of samples on label $j$.}

\begin{table*}[t]
\centering
\setlength{\tabcolsep}{3.2mm}{
\begin{threeparttable}
\caption{{The results of top 10 models according to the sample-based metrics}}
\label{tab:basemodel}
\centering
\begin{tabular}{|c|cccccccccc|}
\hline
Model id & 1 & 2 & 3 & 4 & 5 & 6 & 7 & 8 & 9 & 10 \\ \hline\hline
\begin{tabular}[c]{@{}c@{}}MLC\\ algorithm\end{tabular} & CDN & CDN & BR & CDN & RT & PS & RAkELd & CC & CC & CT \\\hline
\begin{tabular}[c]{@{}c@{}}basic\\ classifier\end{tabular} & J48 & LMT & \begin{tabular}[c]{@{}c@{}}Random\\ Forest\end{tabular} & REPTree & LMT & \begin{tabular}[c]{@{}c@{}}Random\\ Forest\end{tabular} & \begin{tabular}[c]{@{}c@{}}Random\\ Forest\end{tabular} & \begin{tabular}[c]{@{}c@{}}Random\\ Forest\end{tabular} & REPTree & \begin{tabular}[c]{@{}c@{}}Random\\ Forest\end{tabular} \\\hline\hline
hamming loss & 0.133 & 0.128 & 0.139 & 0.133 & 0.161 & {\textbf{0.122}} & 0.133 & 0.156 & 0.133 & 0.156 \\
zero-one loss & 0.533 & 0.467 & 0.533 & 0.500 & 0.533 &{\textbf{0.433}}  & 0.500 & 0.533 & 0.533 & 0.533 \\
F1-score & 0.696 & 0.723 & 0.702 & 0.704 & 0.707 & {\textbf{0.741}} & 0.719 & 0.702 & 0.693 & 0.702 \\
sample-ACC & \textbf{0.733} & \cellcolor{graytbl}\textbf{0.733} & 0.700 & 0.700 & 0.700 & 0.700 & 0.700 & 0.667 & 0.667 & 0.667\\\hline
\end{tabular}
\begin{tablenotes}
\scriptsize
\item *To make the table more readable, we highlight the locally optimal results (for sample-based metrics) with bold font and the key determinant(s) of the overall best model with gray color. (also for Table~\ref{table:authenticity})
\end{tablenotes}
\end{threeparttable}
}
\end{table*}

\begin{table*}[t]
\centering
\setlength{\tabcolsep}{3.45mm}{
\begin{threeparttable}
\caption{{The label-based evaluation results of top 10 models}}
\label{tab:label_basemodel}
\begin{tabular}{|c|cccccccccc|}
\hline
Model id & 1 & 2 & 3 & 4 & 5 & 6 & 7 & 8 & 9 & 10 \\ \hline\hline
\begin{tabular}[c]{@{}c@{}}MLC\\ algorithm\end{tabular} & CDN & CDN & BR & CDN & RT & PS & RAkELd & CC & CC & CT \\\hline
\begin{tabular}[c]{@{}c@{}}basic\\ classifier\end{tabular} & J48 & LMT & \begin{tabular}[c]{@{}c@{}}Random\\ Forest\end{tabular} & REPTree & LMT & \begin{tabular}[c]{@{}c@{}}Random\\ Forest\end{tabular} & \begin{tabular}[c]{@{}c@{}}Random\\ Forest\end{tabular} & \begin{tabular}[c]{@{}c@{}}Random\\ Forest\end{tabular} & REPTree & \begin{tabular}[c]{@{}c@{}}Random\\ Forest\end{tabular} \\\hline\hline
L1 & 0.833 & \textbf{0.867} & 0.833 & 0.800 & 0.833 & 0.833 & 0.833 & 0.767 & 0.800 & 0.767 \\
L2 & 0.800 & 0.800 & \textbf{0.833} & \textbf{0.833} & 0.800 & 0.800 & 0.800 & 0.800 & \textbf{0.833} & 0.800 \\
L3 & \textbf{0.933} & \textbf{0.933} & 0.867 & \textbf{0.933} & 0.867 & \textbf{0.933} & \textbf{0.933} & 0.867 & \textbf{0.933} & 0.867 \\
L4 & 0.933 & 0.900 & 0.933 & 0.933 & 0.867 & \textbf{0.967} & \textbf{0.967} & 0.933 & 0.933 & 0.933 \\
L5 & \textbf{0.900} & \textbf{0.900} & \textbf{0.900} & \textbf{0.900} & \textbf{0.900} & \textbf{0.900} & \textbf{0.900} & \textbf{0.900} & \textbf{0.900} & \textbf{0.900} \\
L6 & 0.800 & \textbf{0.833} & 0.800 & 0.800 & 0.767 & \textbf{0.833} & 0.767 & 0.800 & 0.800 & 0.800\\\hline
L-avg. & {0.867} & \cellcolor{graytbl}\textbf{0.872} & 0.861 & 0.867 & 0.839 &0.867 & 0.867 &  0.845& 0.867&0.845\\
\hline
\end{tabular}
\begin{tablenotes}
\scriptsize
\item *L1-6 refers to the label-ACC of 6 predefined labels (i.e., ``SMS-related'', ``lock in'', ``Re-infection'', ``Telephony-related'', ``Ads'', ``Internet-related'') and L-avg. is the average of all L1-L6.
\item *To make the table more readable, we highlight the locally optimal results (for the label-ACC of each label) with bold font and the key determinant (L-avg.) of the overall best model with gray color.
\end{tablenotes}
\end{threeparttable}
}
\end{table*}

\subsection{{RQ1: Which combination of MLC and BC algorithms is best for multi-label classification of malware?}}
\label{evaluation:ML_MLC_model}
{In this experiment, to find out the best combinations of multi-label classification and basic classification algorithms, we evaluate the effectiveness of diverse machine learning models on the metrics proposed in \S~\ref{evaluation:metrics}.}

\subsubsection{Dataset}
\label{evaluation:ML_MLC_model:dataset}
{As shown in Table~\ref{tab: sample number of each class}, we use the dataset of 180 labeled malware samples as the ground truth. Due to the limited size of the dataset, using a common data split ratio (i.e., 8:2) like other approaches which perform on large-scale datasets can seriously undermine the validity of the evaluation result. Especially, for the labels with a very small number of samples, such as ``Telephony-related'', if the test set has only 2 samples, it will be quite hard to judge the actual classification ability since there are only 3 possible test results (i.e., 0\%, 50\% or 100\%). Besides, because of the imbalanced data distribution, samples under this label may not exist in the test set, if we adopt a global random data partitioning as usual.
Hence, when dividing the dataset, we ensure that the proportions of samples under each label in the training set and test set are the same.
We split the 180 labeled dataset into a training set of 150 samples and a test set of 30 samples.}

\subsubsection{Experiment setup}
\label{evaluation:ML_MLC_model:setup}
{To evaluate the effectiveness of different multi-label classification (MLC) and basic classification (BC) algorithms, we adopt 70 different combinations of MLC and BC algorithms and train each with the same data configuration. By comparing the results by calculating the selected sample-based and label-based evaluation metrics, we then determine the most suitable combinations for our task.}

\subsubsection{Results}
{We demonstrate the top 10 results from 2 aspects referring to sample-based and label-based metrics, respectively. The full results could be found on the website~\cite{googleSites}.}

% Sample-based
\paragraph{\textbf{Sample-based evaluation}}
{As shown in Table~\ref{tab:basemodel}, the sample-based results of the top 10 models are sorted by the sample-ACC in descending order.
% To make the table more readable, we highlight the locally optimal results (for each metric) with bold font and the key determinant(s) of the overall best model with gray color.
Overall, model~\#1 and \#2 outperform other candidates at 0.733 on the sample-ACC. Between them, model~\#2 using CDN as MLC algorithm and LMT as BC algorithm achieves a better result in terms of higher sample-ACC and F1-score, which are 0.733 and 0.723, and lower hamming loss and zero-one loss, which are 0.128 and 0.467, respectively. This evaluation result reveals that the above combination is able to predict as many malicious behaviors as possible and is less likely to make false classifications. 
Besides the best 2 combinations, model~\#3-7 have the second highest sample-ACC at 0.7, model~\#8-10 have the third highest sample-ACC at 0.667. Due to the limitation of article length and weaker importance of results, the evaluation results of {the} rest 60 combinations are released on the website~\cite{googleSites}}
%other metrics
{We can also notice that model~\#6, which uses PS as MLC algorithm and RandomForest as the BC algorithm, outperforms among all models in terms of hamming loss, zero-one loss, and F1-score at 0.122, 0.433, and 0.741, respectively. That {reveals} the prediction results given by this combination are relatively more comprehensive, {and it will make fewer false classifications.}}
%all
{Through this evaluation, CDN obviously shows their stronger effectiveness in our task since 3 of the top 10 models are using CDN as their MLC algorithm.}

%label-based
\paragraph{\textbf{Label-based evaluation}}
{To further compare the effectiveness of different algorithm combinations towards each specific malicious behavior, we further evaluate the accuracy of the top 10 models on each predefined label.
As shown in Table~\ref{tab:label_basemodel}, we use L1-6 to denote the label-ACC of 6 types of malicious behaviors, which are ``{SMS-related}'', ``{Internet-related}'', ``{Telephony-related}'', ``{Lock-in}'', ``{Ads}'' and ``{Re-infection}'' in order, and use L-avg. to denote the averaged label-ACC (a.k.a., label-based accuracy).
From the results, we can see that all MLC models have the same L5 (a.k.a., the label-ACC of the 5th malicious behavior, namely ``Ads'') at 0.9. We notice that model~\#2, which has the best result in the sample-based evaluation, also outperforms others in the label-based evaluation. The L-avg. of model~\#2 is 0.872. In detail, it reaches the best results on L1,3,5,6 at 0.867, 0.933, 0.9, and 0.833, respectively. These results reveal that the combination of algorithms CDN and LMT has higher confidence in these 4 labels in our task.
Besides, sorted by L-avg. in descending order, the rest are model~\#{1,4,6,7,9, \#3, \#8,10 and \#5 at 0.867, 0.861, 0.845, and 0.839. Among them, model~\#3,4 achieve the highest L2 at 0.833, and model~\#6,7 achieve the highest L4 at 0.967.} }

\begin{table*}[t]
\centering
\setlength{\tabcolsep}{3.1mm}{
\begin{threeparttable}
\caption{The results of Detection-Training with top 10 algorithm combinations on the DREBIN dataset}
\label{tab:drebin:detection_training}
\centering
\begin{tabular}{|c|cccccccccc|}
\hline
model id & 1 & 2 & 3 & 4 & 5 & 6 & 7 & 8 & 9 & 10 \\ \hline\hline
\begin{tabular}[c]{@{}c@{}}MLC\\ algorithm\end{tabular} & CDN & CDN & BR & CDN & RT & PS & RAkELd & CC & CC & CT \\ \hline
\begin{tabular}[c]{@{}c@{}}basic \\ clsssifier\end{tabular} & J48 & LMT & \begin{tabular}[c]{@{}c@{}}Random\\ Forest\end{tabular} & REPTree & LMT & \begin{tabular}[c]{@{}c@{}}Random\\ Forest\end{tabular} & \begin{tabular}[c]{@{}c@{}}Random\\ Forest\end{tabular} & \begin{tabular}[c]{@{}c@{}}Random\\ Forest\end{tabular} & REPTree & \begin{tabular}[c]{@{}c@{}}Random\\ Forest\end{tabular} \\ \hline\hline
\# of samples & \textbf{4,872} & 312 & 16 & \textbf{5,456} & 472 & \textbf{5,456} & 16 & \textbf{1,720} & 72 & \textbf{3,440} \\
hamming loss & \cellcolor{graytbl}\textbf{0.106}&{0.122}  & 0.111 & 0.144& 0.172  & 0.139&0.122  & 0.128 &  0.128& 0.122 \\
zero-one loss &\cellcolor{graytbl}\textbf{0.467} &  0.500& 0.433 &0.500 & 0.533 & 0.533 & 0.467 & \textbf{0.467} & 0.500 &\textbf{0.467} \\
F1-score & \cellcolor{graytbl}\textbf{0.779} & {0.742} &  0.758&0.729 & 0.696 & 0.730 & 0.748 & 0.752 & 0.734 & 0.769 \\
sample-ACC & \cellcolor{graytbl}\textbf{0.867} & 0.833 & 0.767 & 0.767 & 0.800 & 0.733 & 0.767 & 0.767 & 0.800 & 0.767 \\
$\Delta$sample-ACC & \cellcolor{graytbl}\textbf{0.134} & 0.100 & 0.067 & 0.067 & 0.100 & 0.033 & 0.067 & 0.100 & 0.133 & 0.100 \\
Avg. label-ACC &\cellcolor{graytbl}\textbf{0.894}&  0.878&0.889  & 0.856 &  0.828& 0.861 & 0.878 & 0.872 &  0.872& 0.878 \\ \hline
\end{tabular}
\begin{tablenotes}
\scriptsize
\item *To make the table more readable, we highlight the locally optimal results (only if \# of samples $>$ 1k, the model is considered; best one(s) for other metrics) with bold font and the key determinants of the overall best models with gray color. (also for Table~\ref{tab:virushare:detection_training}, \ref{tab:result-N-DREBIN} and \ref{tab:result-N-VirusShare})
\end{tablenotes}
\end{threeparttable}
}
\end{table*}

\begin{figure*}[ht]
	\centering  
	\subfigcapskip=0pt 
	\subfigure[CDN+J48 (model~\#1)]{
	    \label{Fig:drebin:CDN+J48}
		\includegraphics[width=0.33\linewidth]{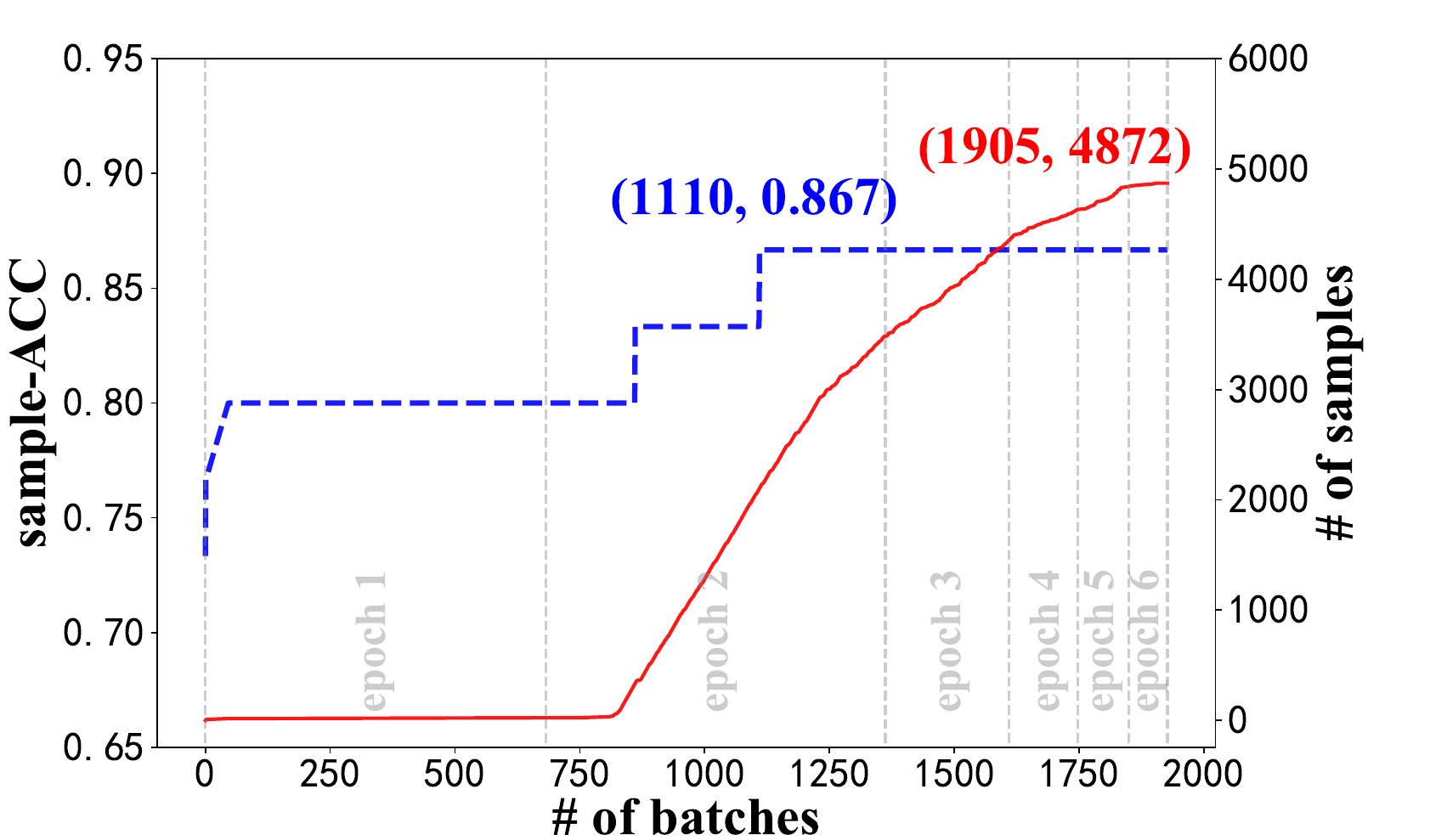}}\hspace{-3mm}
	\subfigure[CDN+REPTree (model~\#4)]{
	    \label{Fig:drebin:CDN+REPTree}
		\includegraphics[width=0.33\linewidth]{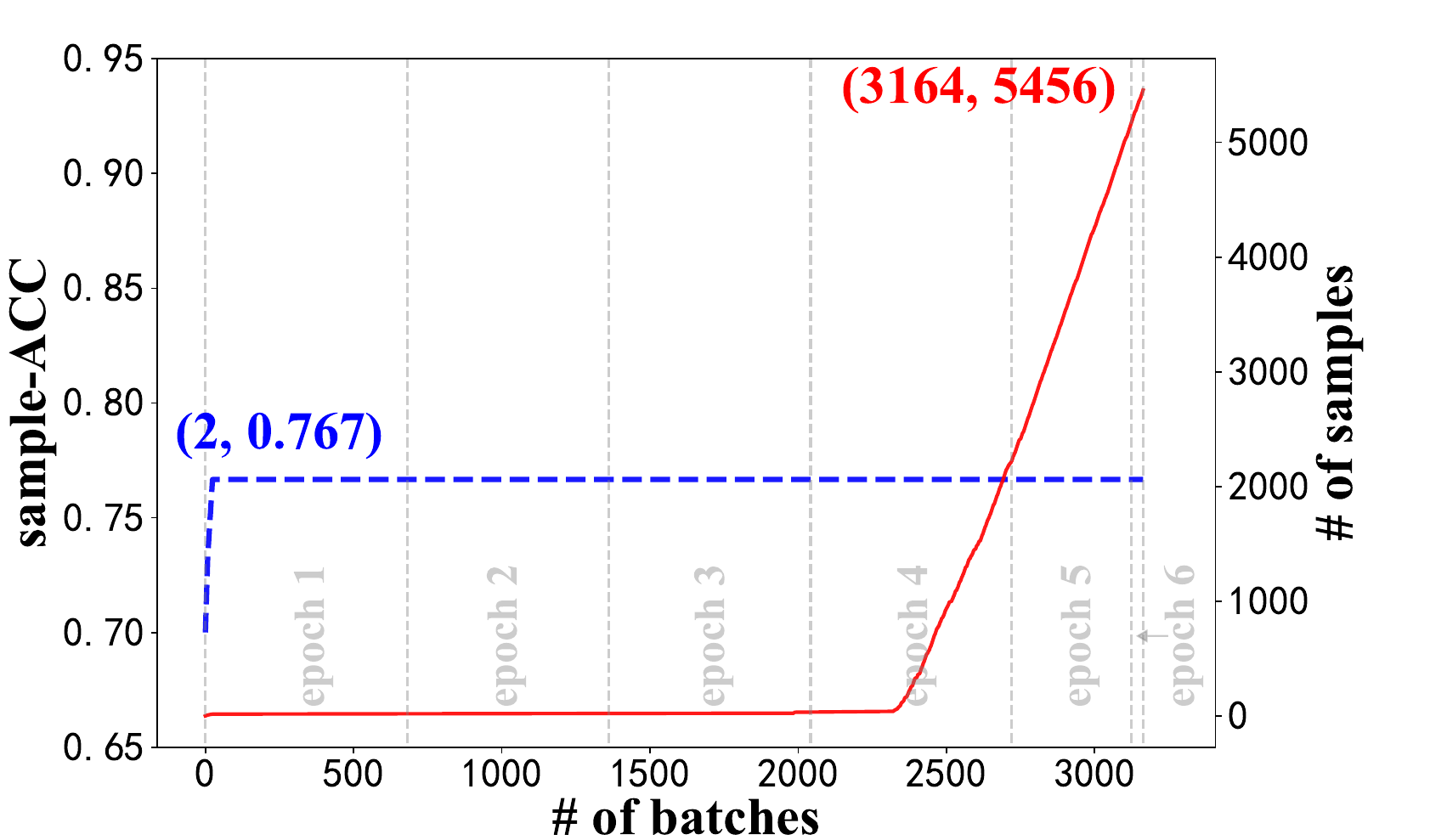}}\hspace{-3mm}
	\subfigure[PS+RandomForest (model~\#6)]{
	    \label{Fig:drebin:PS+RandomForest}
		\includegraphics[width=0.33\linewidth]{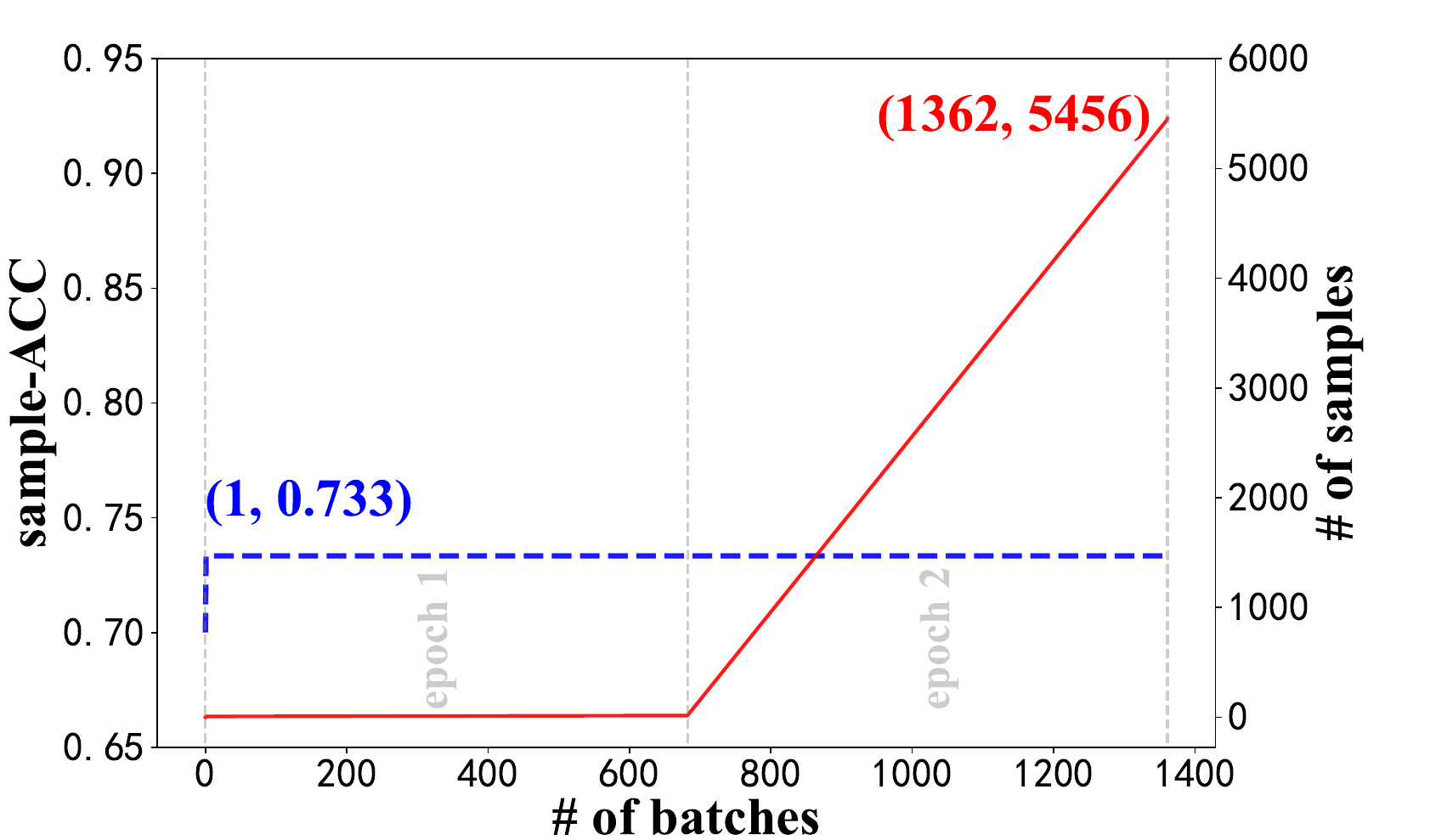}}\hspace{-3mm}
	\subfigure[CC+RandomForest (model~\#8)]{
		 \label{Fig:drebin:CC+RandomForest}
		\includegraphics[width=0.33\textwidth]{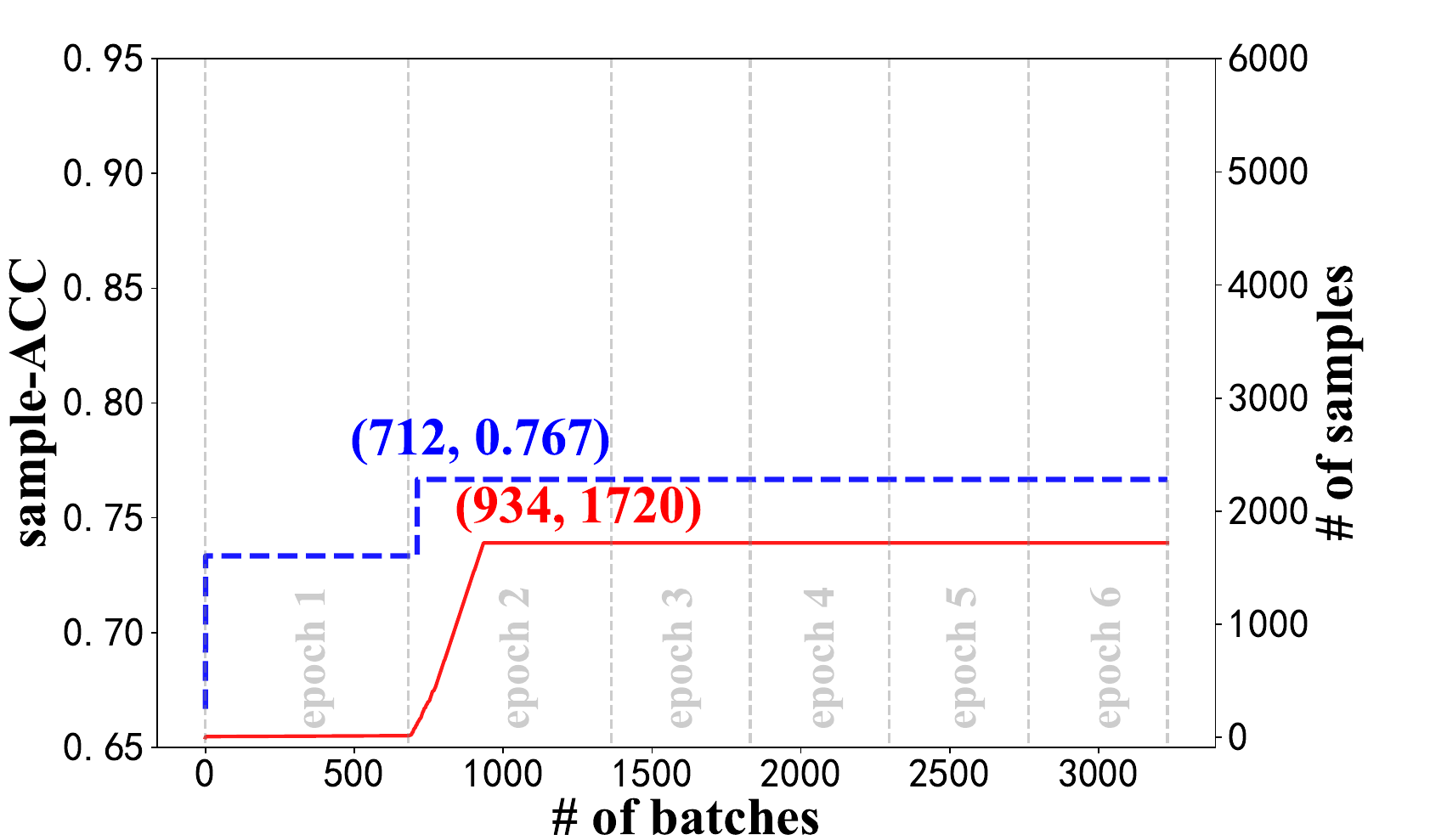}}\hspace{-3mm}
	\subfigure[CT+RandomForest (model~\#10)]{
		  \label{Fig:drebin:CT+RandomForest}
	\includegraphics[width=0.33\textwidth]{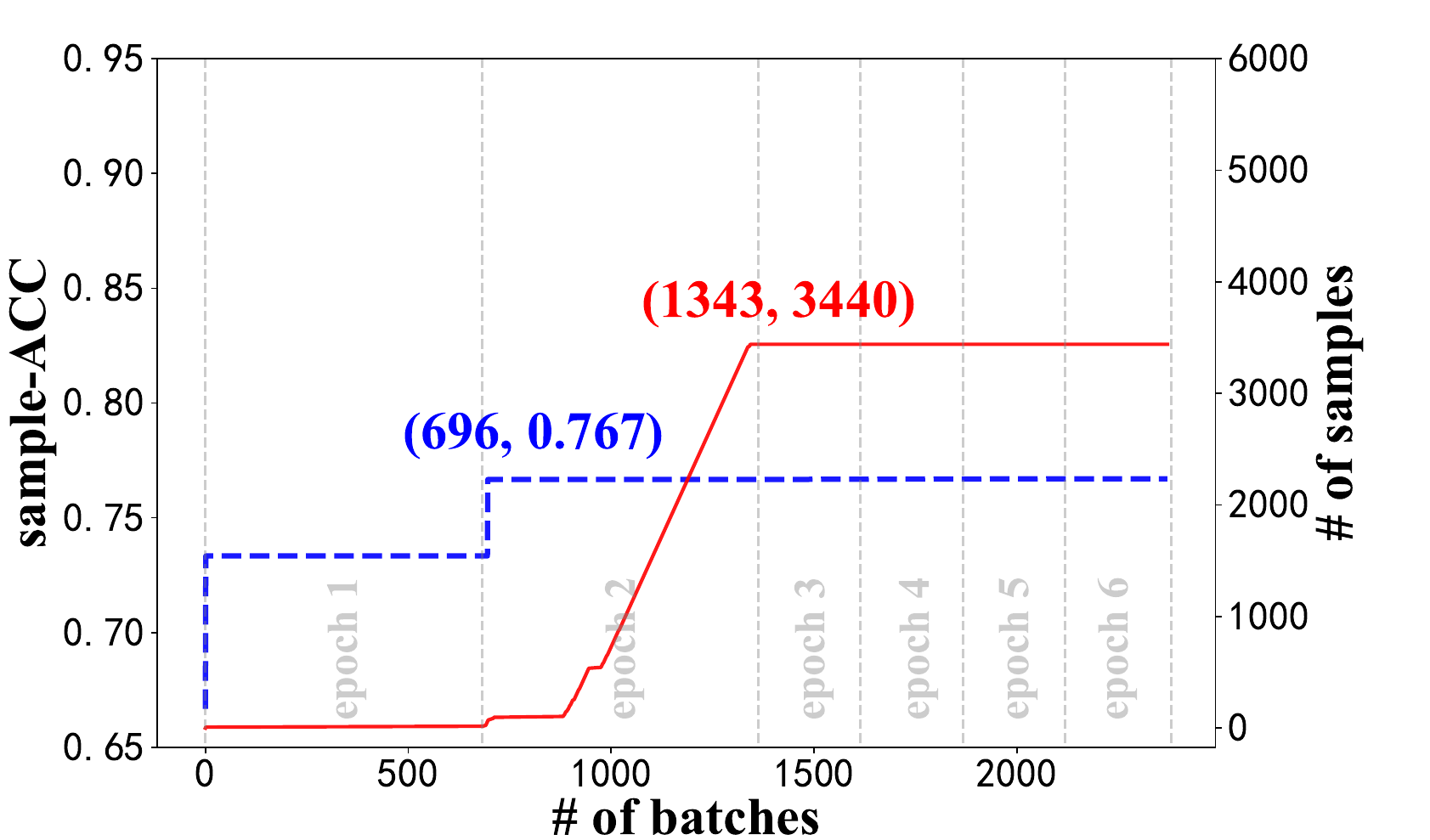}}
	\includegraphics[width=0.2\linewidth]{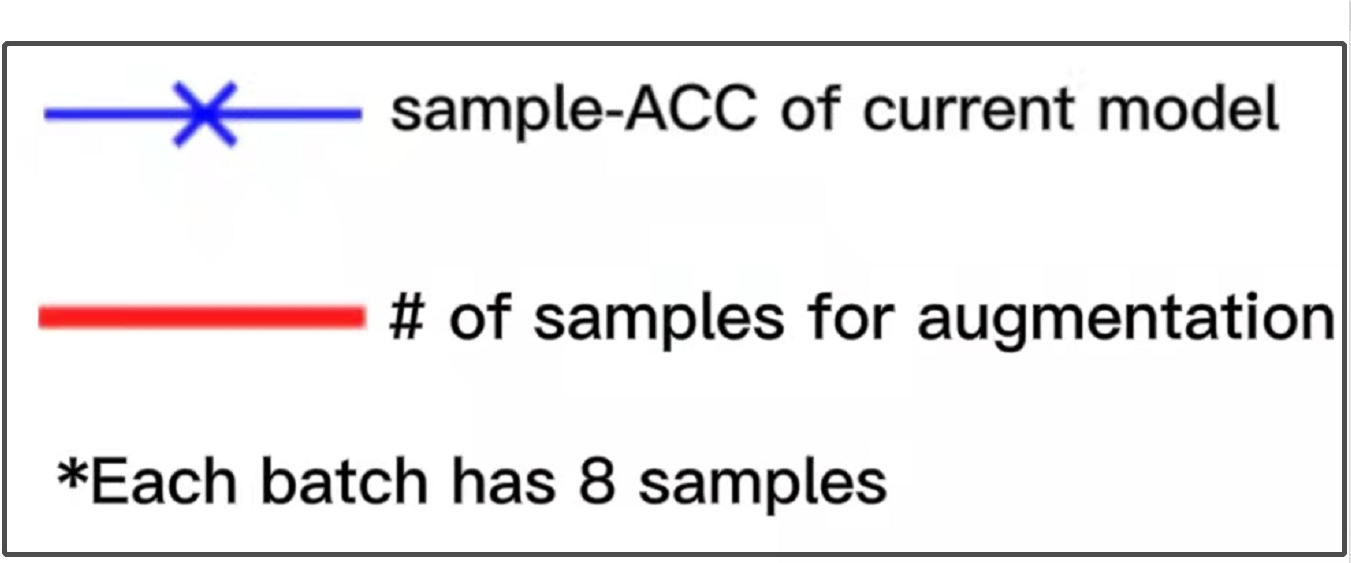}\hspace{-3mm}
		\\
	\caption{The plots of accuracy improvement and data augmentation on the DREBIN dataset}\label{Fig:process of data augmentation-drebin}
\end{figure*}

\begin{tcolorbox}[size=title,opacityfill=0.1,breakable]
\noindent\textit{\textbf{{Conclusion}: }{By training with 70 different algorithm combinations, we basically confirm the candidate MLC and BC algorithms that could achieve better results for our task. The best practice reaches 0.733 and 0.872 on sample-ACC and averaged label-ACC.}}
\end{tcolorbox}

\subsection{{RQ2: Can Detection-Training effectively augment the dataset? What is the best batch-size for enhancement?}}
\label{evaluation:detection_training}
{In this section, we investigate the effectiveness of Detection-Training from 2 aspects. First, we conduct an experiment to see which ML algorithm can perform well with Detection-Training. Second, we further tune the parameters defined in Detection-Training to achieve a better result on accuracy improvement and data augmentation.}

\subsubsection{Capability Evaluation of Detection-Training on Accuracy Improvement and Data Augmentation}
\label{evaluation:detection_training:active_learning}
{To find out the most suitable algorithm combinations in our task, we conduct the experiment that trains 10 combinations on 2 different malware datasets. We evaluate the capability according to 2 factors, such as the \# of samples for data augmentation and the accuracy improvement of the MLC model.}

\paragraph{\textbf{Evaluation on the DREBIN datset}}

{\textbf{Dataset.}}
\label{evaluation:detection_training:active_learning:dataset_drebin}
{In this experiment, we adopt a setting on the labeled dataset as same as \S~\ref{evaluation:ML_MLC_model:dataset}, because it is necessary to keep the test set remaining unchanged while comparing the accuracy between the base model and better models, which obtained through Detection-Training. Besides, two more datasets without manual labeling are used as the ``Malware for data augmentation'' in the Detection-Training module which is introduced in \S~\ref{approach:detection_training}. The first one is a widely used public dataset, namely DREBIN. As introduced in \S~\ref{evaluation:dataset:drebin}, the malware samples in DREBIN were collected in the period of 2010 to 2012. During the past decade, the attack principle of malware also changes with the improvement of the Android system mechanism and implementation. Hence, the evaluation results on DREBIN may be affected by the evolution of the malicious behaviors implemented in malware. 
}

\noindent{\textbf{Experiment setup.}}
\label{evaluation:detection_training:active_learning:setup}
{To evaluate the capability of the proposed active learning framework on the accuracy improvement and data augmentation by using the results of base models in \S~\ref{evaluation:ML_MLC_model} as the baseline, we adopt 10 algorithm combinations as same as those of the top 10 best base models. In the experiment, we follow the steps described in \S~\ref{approach:detection_training}. We set 8 as the default batch size, which means 8 samples from the ``data pool'' are fed into the ``base/updated model'' in each round of workflow, and 6 as the default epoch number, which means traversing ``data pool'' for 6 times by adding the shuffled the unused data in step~\step{5} at the end of each epoch.}

\noindent{\textbf{Result on the DREBIN dataset.}}
\label{evaluation:detection_training:active_learning:drebin}
{Table~\ref{tab:drebin:detection_training} presents the evaluation results of improving the classification accuracy and size of the usable dataset by applying Detection-Training to the DREBIN dataset. Overall, the accuracy increase (i.e., $\Delta$sample-ACC) can reach up to 0.134 on the test set by comparing to the base models, even the worst case has an increase at 0.033. Among the 10 selected algorithm combinations, model~\#1, which uses CDN as MLC algorithm and J48 as BC algorithm, achieves the best result, which shows a significant improvement on all evaluation metrics with a successful data augmentation. Specifically, model~\#1 reaches the lowest hamming and zero-one loss at 0.106 and 0.467, the highest F1-score, sample-ACC, and averaged label-ACC at 0.779, 0.867, and 0.894, respectively, and the most successful accuracy improvement at 0.134 based on a data augmentation with 4,872 malware samples. Besides, we notice that there are 4 more combinations of MLC and BC algorithms, which are used in constructing model~\#4, 6, 8, 10, and can also expand the dataset with thousands of additional malware samples from DREBIN. Among them, model~\#4 and \#6 have increased the training set to the largest size with 5,456 samples. Model~\#4 improves the sample-ACC to 0.733, which is 0.033 higher than that of the base model with the same algorithm combination. The averaged label-ACC of it is 0.856, which has a little decrease by comparing to the result of its base model (0.867). Model~\#6 also improves the sample-ACC to 0.733, which is 0.033 higher than that of the base model with the same algorithm combination.}

{To investigate the precise improvement in both accuracy and data augmentation during the entire active learning progress, we also plot the detailed results in Fig.~\ref{Fig:process of data augmentation-drebin}. Generally, the plots present the run-time sample-ACC and the total number of involved samples by augmentation at each batch. We pick 6 as the epoch number in the experiment, because we observe the accuracy and number of samples always remaining unchanged after 6 epochs. {Note that each epoch accumulates batches based on the previous epoch, the number of batches is not set to zero after one epoch ends, but it still keeps increasing.} Totally, there are 5 models (i.e., model~\#1, 4, 6, 8, and 10) that obtained a data augmentation of more than one thousand samples. Fig.~\ref{Fig:drebin:CDN+J48} shows the result of model~\#1, which uses CDN and J48 as MLC and BC algorithms. We can see that in the early stage of Detection-Training, along with the increasing \# of batches, the sample-ACC and \# of augmented samples keep increasing. After the \# of batches reaches 1,110, the sample-ACC remains unchanged at 0.876. And, the \# of samples for augmentation is still able to increase. In epoch 6, when 1,905 batches have been fed to the base/update models, the \# of samples reaches 4,872. After that, the \# of augmented data hardly ever grows significantly.
As shown in Fig.~\ref{Fig:drebin:CC+RandomForest} and Fig.~\ref{Fig:drebin:CT+RandomForest}, model~\#8 and \#10, which were trained with the combinations of CC and RandomForest and CT and RandomForest, also have a similar tendency.
For these above 3 combinations, there is always such a situation observed, when the batches of involved samples reach a certain number, the classification capability of the model and \# of samples for augmentation can no longer increase. For model~\#8 and \#10, the lift of the curves stops in epoch 2.
For model~\#4, when the \# of batches reaches 2, the sample-ACC of the current model reaches the peak at 0.767. In epoch 4, \# of samples for augmentation starts to increase significantly, and in epoch 6, when \# of batches equals to 3,164, all 5,456 samples in the DREBIN dataset have been labeled with pseudo labels and added to the training set. A similar situation happens to model~\#6, which uses the combinations of PS and RandomForest. Fig~\ref{Fig:drebin:PS+RandomForest} shows that in epoch 2, it also adds all 5,456 samples from the DREBIN dataset to the training set.}

\begin{figure}[t]
\centering
\includegraphics[width=1\linewidth]{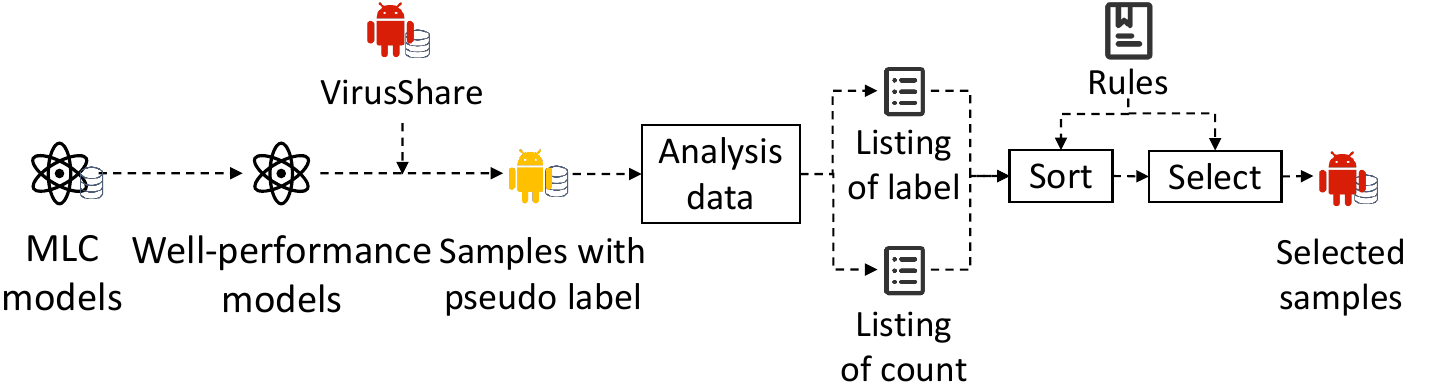}
\caption{The process of selecting samples from VirusShare dataset}
\label{fig:VirusShare-building-2}
\end{figure}

\begin{table*}[ht]
\caption{The results of Detection-Training with top 10 algorithm combinations on VirusShare dataset}
\label{tab:virushare:detection_training}
\centering
\setlength{\tabcolsep}{3.05mm}{
\begin{tabular}{|c|cccccccccc|}
\hline
model id & 1 & 2 & 3 & 4 & 5 & 6 & 7 & 8 & 9 & 10 \\ \hline\hline
\begin{tabular}[c]{@{}c@{}}MLC\\ algorithm\end{tabular} & CDN & CDN & BR & CDN & RT & PS & RAkELd & CC & CC & CT \\ \hline
\begin{tabular}[c]{@{}c@{}}basic\\ clsssifier\end{tabular} & J48 & LMT & \begin{tabular}[c]{@{}c@{}}Random\\ Forest\end{tabular} & REPTree & LMT & \begin{tabular}[c]{@{}c@{}}Random\\ Forest\end{tabular} & \begin{tabular}[c]{@{}c@{}}Random\\ Forest\end{tabular} & \begin{tabular}[c]{@{}c@{}}Random\\ Forest\end{tabular} & REPTree & \begin{tabular}[c]{@{}c@{}}Random\\ Forest\end{tabular} \\ \hline\hline
\# of samples & \textbf{5416} & 536 & 72 & 120 & 960 & \textbf{5496} & \textbf{2432} & \textbf{1088} & 24 & \textbf{1848} \\
hamming loss & 0.133& 0.111 &0.117  & 0.139 & 0.189 & 0.139 &0.144  &\textbf{0.128}  & 0.128 &\textbf{0.128}  \\
zero-one loss & 0.500& 0.433 &0.467  &0.500  &0.600  &0.533  &0.533  &0.500  &0.533  &\textbf{0.467}  \\
F1-score & 0.733 & 0.753 & 0.747 & 0.729 & 0.641 & 0.730 &0.720  & \textbf{0.736} &0.746  &0.719  \\
sample-ACC & \cellcolor{graytbl}\textbf{0.833} & 0.833 & 0.767 & 0.767 & 0.767 & 0.733 & 0.767 & 0.733 & 0.8 & 0.733 \\
$\Delta$sample-ACC & \cellcolor{graytbl}\textbf{0.100} & 0.100 &  0.067&  0.067&  0.067& 0.033 & 0.067 &0.066  &  0.133& 0.066 \\
Avg. label-ACC & 0.867 & 0.889 &0.883  &0.861  &0.811  &0.861  &0.856  &\textbf{0.872}  &0.872  &\textbf{0.872}  \\ \hline
\end{tabular}
}
\end{table*}

\begin{figure*}[ht]
	\centering 
% 	\subfigbottomskip=2pt %两行子图之间的行间距
	\subfigcapskip=0pt 
	\subfigure[CDN+J48 (model~\#1)]{
	    \label{Fig:virusshare:CDN+J48}
		\includegraphics[width=0.33\linewidth]{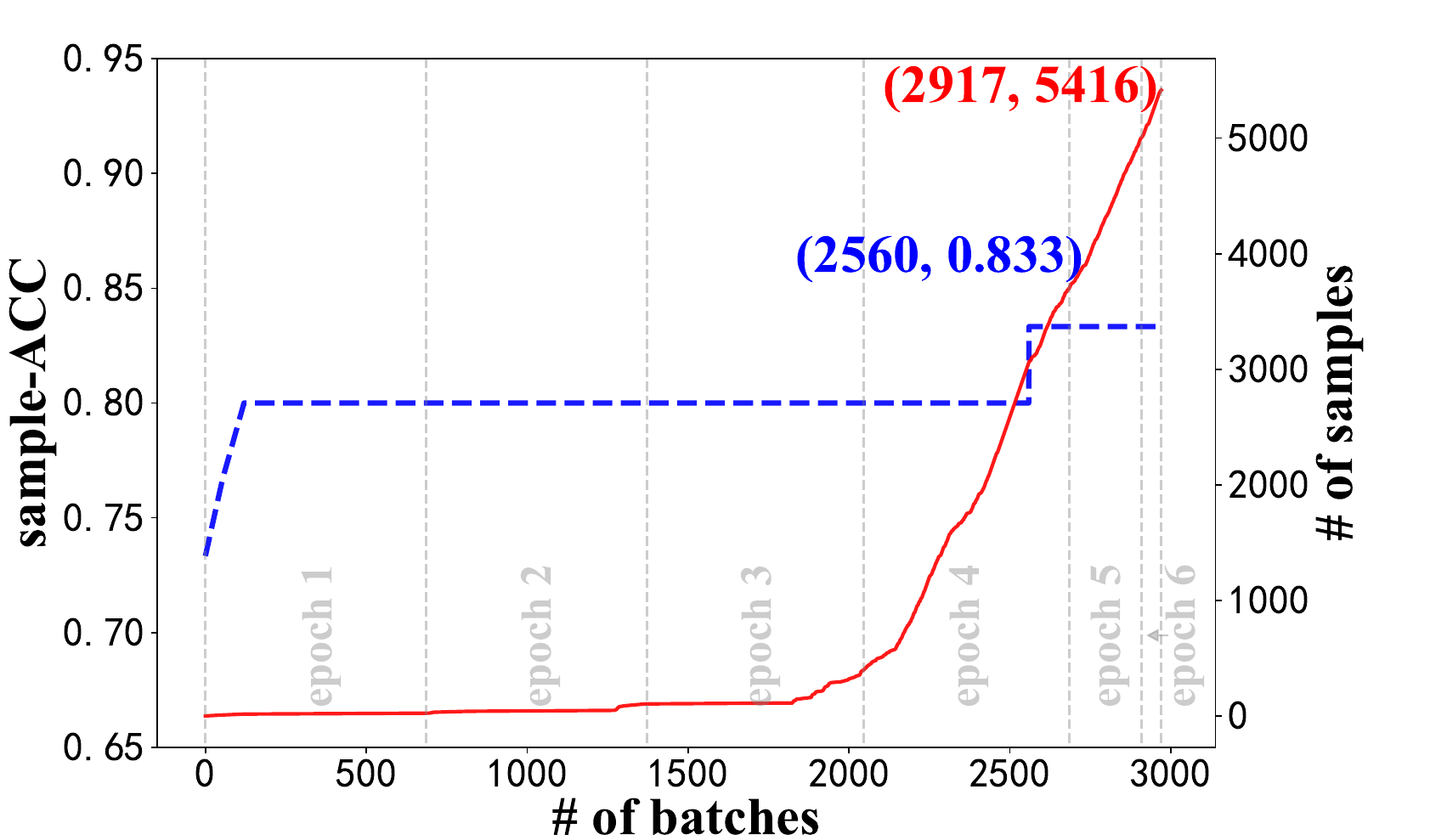}}\hspace{-3mm}
	\subfigure[PS+RandomForest (model~\#6)]{
	    \label{Fig:virusshare:PS+RandomForest}
		\includegraphics[width=0.33\linewidth]{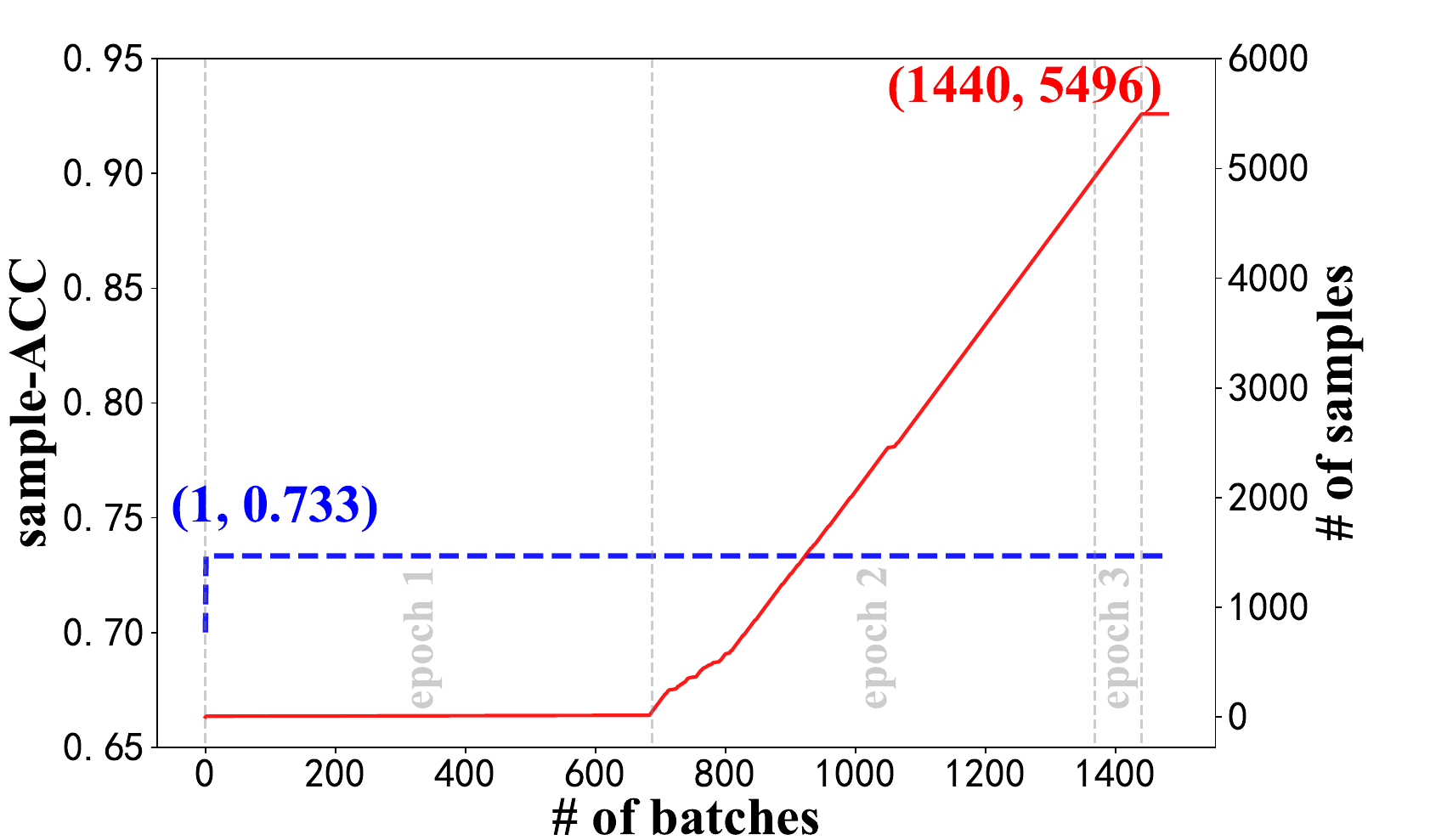}}\hspace{-3mm}
	\subfigure[RAkELd+RandomForest (model~\#7)]{
	    \label{Fig:virusshare:RK+RF}
		\includegraphics[width=0.33\linewidth]{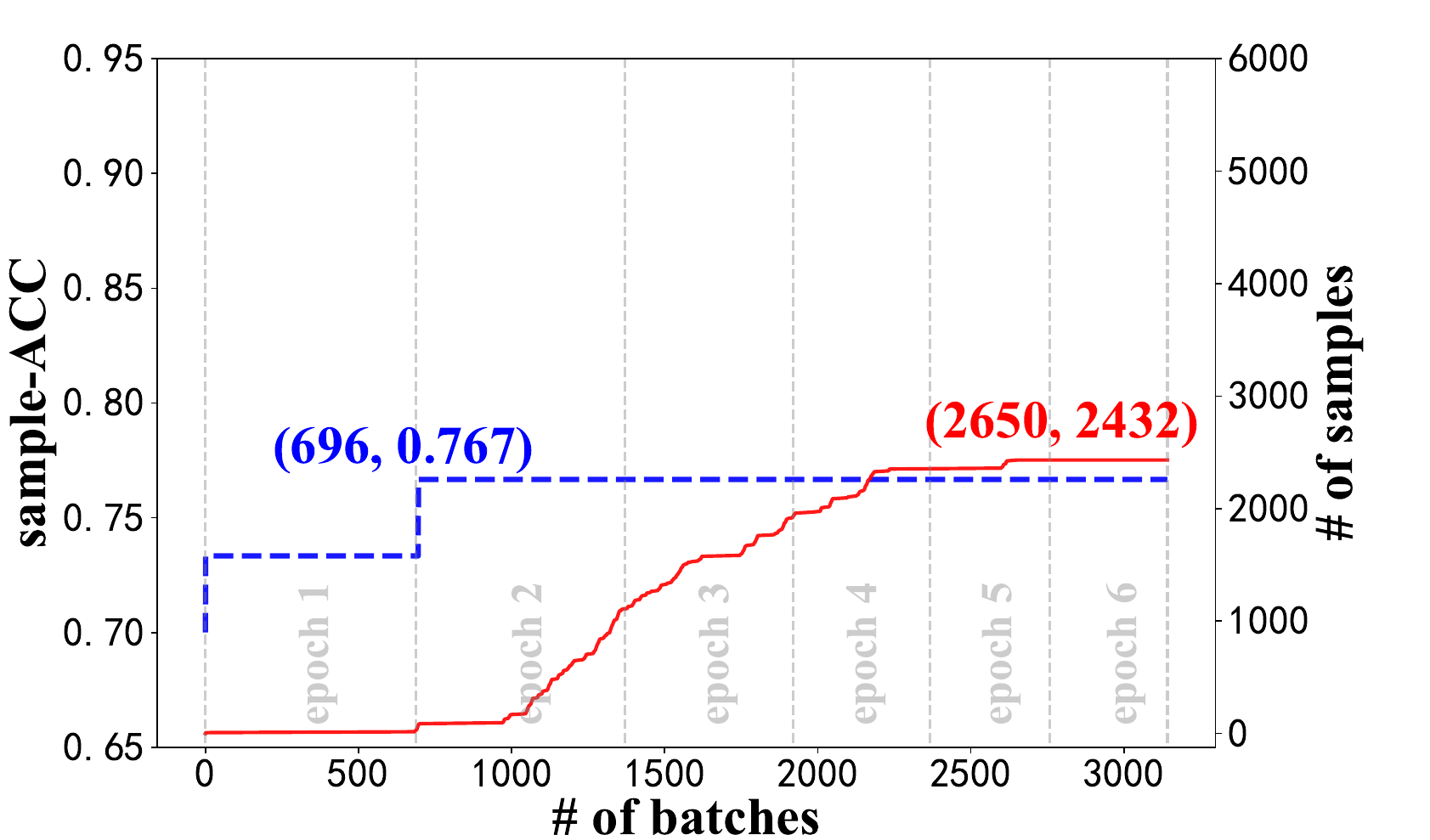}}\hspace{-3mm}
	\subfigure[CC+RandomForest (model~\#8)]{
		 \label{Fig:virusshare:CC+RandomForest}
		\includegraphics[width=0.33\textwidth]{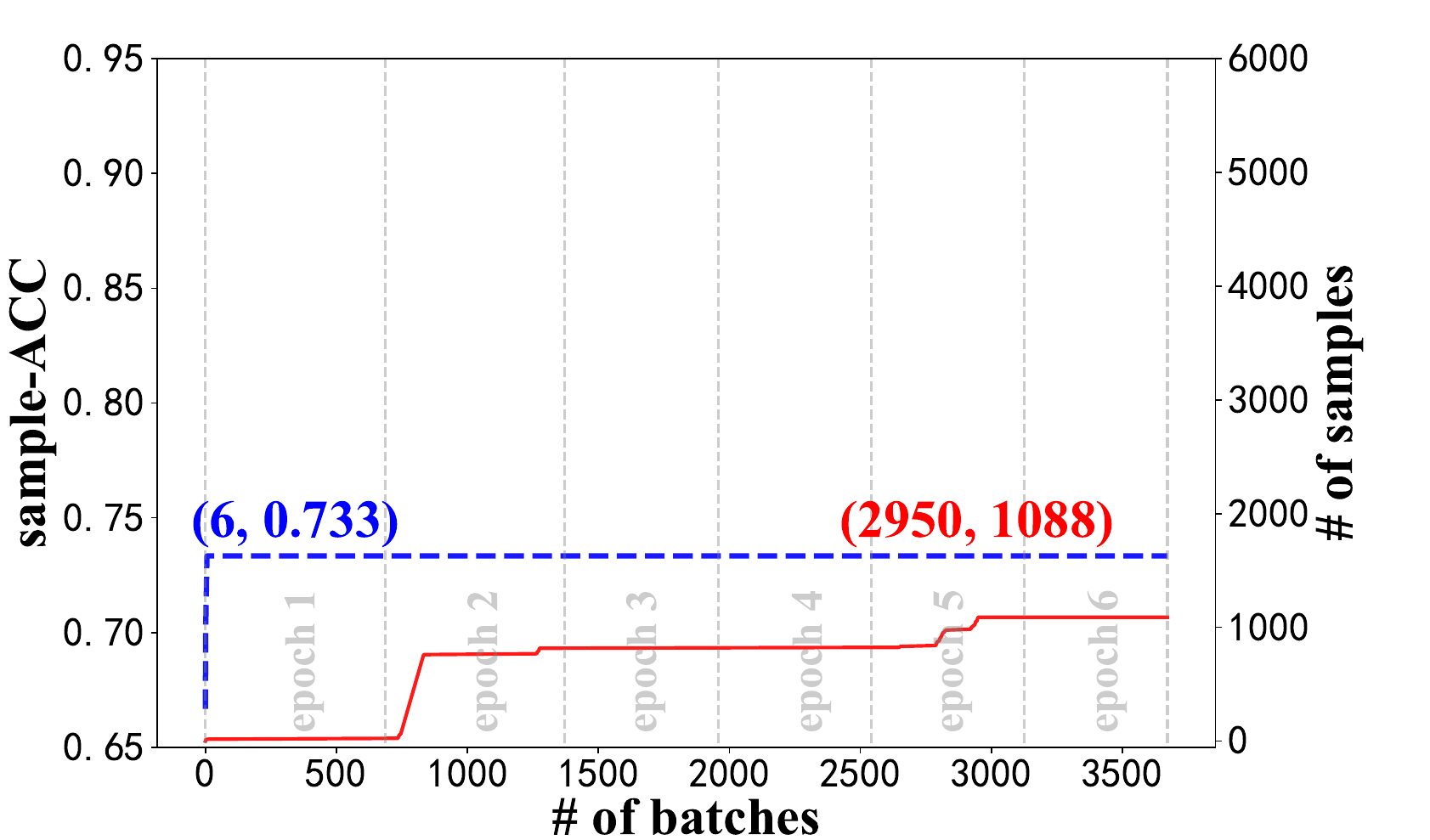}}\hspace{-3mm}
	\subfigure[CT+RandomForest (model~\#10)]{
		  \label{Fig:virusshare:CT+RandomForest}
	\includegraphics[width=0.33\textwidth]{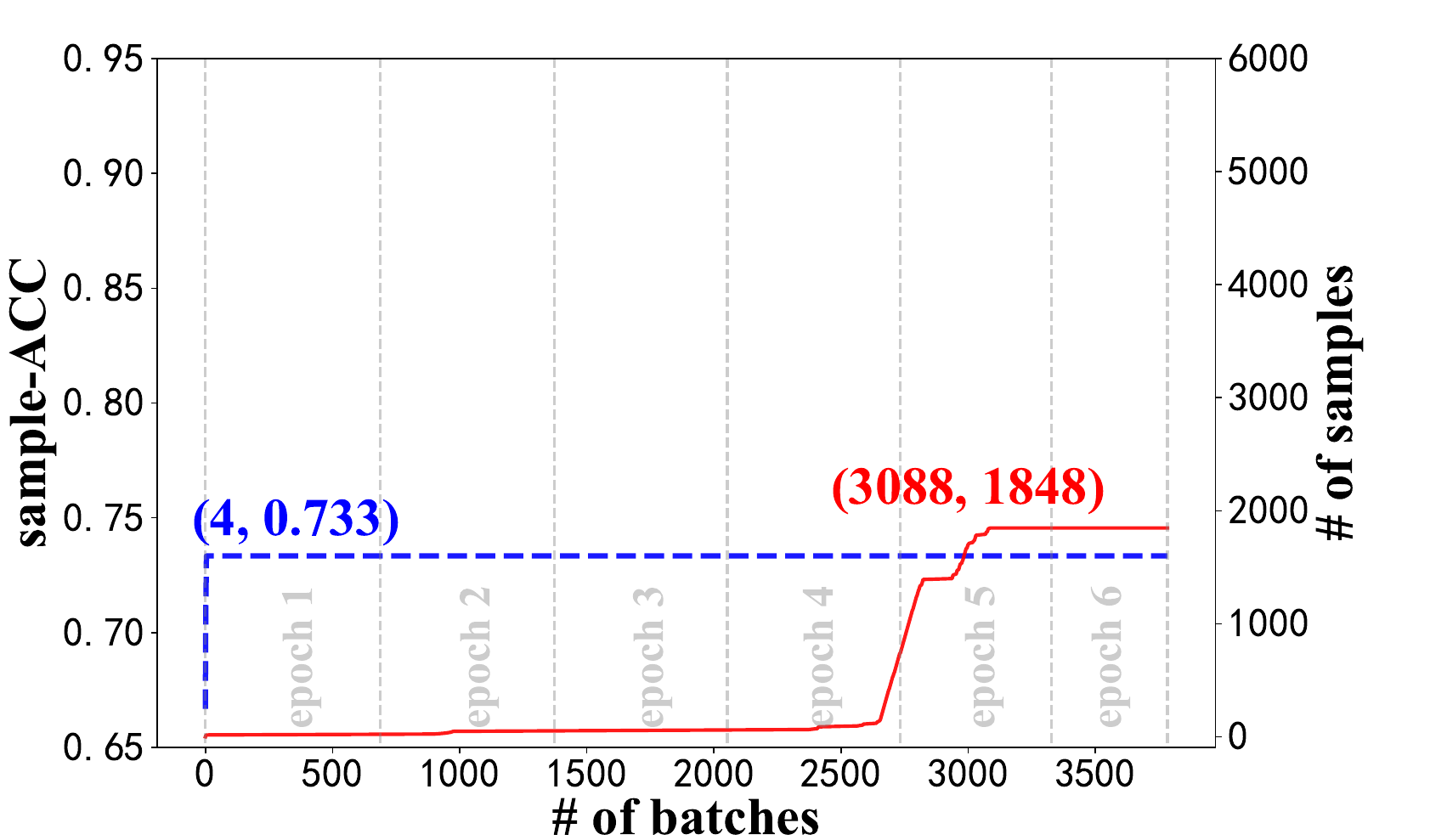}}
	\includegraphics[width=0.2\linewidth]{images/temp/temp.png}\hspace{-3mm}
		\\
	\caption{The plots of accuracy improvement and data augmentation on VirusShare dataset}\label{Fig:process of data augmentation-Virusshare}
\end{figure*}

\paragraph{\textbf{Evaluation on the VirusShare dataset}}\label{data:virusshare}
\textbf{Dataset.}
{VirusShare has tens of thousands of Android malware between 2018 and 2022, the malicious behavior of these samples is varied. In order to balance the data on each label, we propose a method for selecting samples, which selects a subset of samples from VirusShare. The specific data selection process is as shown in Fig.~\ref{fig:VirusShare-building-2}.} 

{First, we select 3 well-performing models from existing MLC models. 
Among them, one model comes from the basic models which are constructed in \S~\ref{evaluation:ML_MLC_model}. As Table~\ref{tab:basemodel} and Table~\ref{tab:label_basemodel} show, the algorithm combination of CDN and LMT (model \#2) achieves the best results. 
The other two models are built in the process of detection-training using the DREBIN dataset in \S~\ref{evaluation:detection_training:active_learning:drebin}, they are trained with the algorithm combinations of CDN and J48 (model \#1), CDN and REPTree (model \#4) shown in Table~\ref{tab:drebin:detection_training}. 
We then use these three well-performing models to predict pseudo-labels for each of the 100,000 samples from VirusShare and get three pseudo-labeled sample datasets. 
Note that, for some same samples, the pseudo-labels predicted by the three models may be different. 
After that, we look into the statistics of each sample referring to these three pseudo-labeled sample datasets and mark each sample with two vectors with lengths at 6, namely \textit{label} and \textit{count}, respectively.
For the vector \textit{label}, the index i represents the number of models whose i-th pseudo-label of the sample is equal to 1.
For the vector \textit{count}, count[j] indicates the number of models that totally predict j labels for this sample. 
After getting the \textit{label} and \textit{count} vectors of all samples, we next sort them and pick samples based on the following 2 rules. Rule \#1 is we want to have at least 1,000 samples on each label. Rule \#2 is it would be better for the selected samples to have as many labels equal to 1 as possible because we want the selected samples to have diverse kinds of behaviors.
We first sort them according to the value of label[i]. 
The larger the label[i] is, the more likely the sample actually has malicious behavior related to this label. For each label, we select the top 1,000 samples. 
Considering the value range of label[i] is [0, 3], all the samples with a high probability of an equal label[i]. 
When such a conflict occurs, we refer to Rule \#2 and perform a secondary analysis on them based on the \textit{count} vector. 
Specifically, we compare the maximum index of each sample whose count[j] is not equal to zero and select a sample with a larger index, which means that it may have more types of malicious behaviors. 
When the indexes of two samples are the same, we compare the value of their count[j] and select the one with a larger count[j], which indicates that it has higher credibility of having actual malicious behavior under label j. 
Through the above steps, we finally select 5,500 samples from VirusShare~\cite{0VirusShare} and construct a relatively balanced VirusShare dataset.}

\noindent\textbf{Experiment setup.}
{In this experiment, we adopt an experiment setup as same as \S~\ref{evaluation:detection_training:active_learning:setup}}

\noindent\textbf{Result on the VirusShare dataset.}
\label{evaluation:detection_training:active_learning:virusshare}
{Table~\ref{tab:virushare:detection_training} presents the evaluation results of Detection-Training with top 10 algorithm combinations on VirusShare dataset. The top results are consistent with the markup of the Table~\ref{tab:drebin:detection_training}. In general, the sample-ACC of these models comparerevise{d} to those of base models can be improved $\ge$ 0.33 on the test set, and the best case has an increase at 0.1. There are totally 5 combinations of MLC and BC algorithms, which are used in constructing model~\#1, 6, 7, 8, 10, and can expand the train set with thousands of additional malware samples from the VirusShare dataset. Among all of them, model~\#1, which uses the algorithm combination of CDN and J48, achieves the best result. Its sample-ACC is 0.833, which is 0.1 more than that of its base model. Model \#1 also enlarge the training set to the second largest size at 5,416. Besides, model~\#6 has increased the training set to the largest size with 5,496 samples and improved the sample-ACC to 0.733, which is 0.033 higher than that of its base model. The averaged label-ACC of it is 0.861, which has a little decrease by comparing to that of its base model (0.867).}

{We also plot the detailed results of model~\#1, 6, 7, 8, 10 in Fig.~\ref{Fig:process of data augmentation-Virusshare}. The x-axis represents \# of batches used in Detection-Training. Generally, the plots present the run-time sample-ACC and the total number of involved samples by augmentation at each batch. As a default, we set the epoch number to 6.
In Fig.~\ref{Fig:virusshare:CDN+J48} which shows the result of the combination of CDN and J48 (model~\#1), along with the increasing \# of batches, the sample-ACC and \# of augmented samples keep increasing at the early stage of Detection-Training. when the \# of batches reaches 2,560, the sample-ACC reaches a peak at 0.833. And after that, the \# of samples for augmentation is still able to increase. In epoch 6, when 2,917 batches have been fed into the base/update models, the \# of samples reaches 5,416 and the \# of augmented data hardly ever grows significantly after that.
As shown in Fig.~\ref{Fig:virusshare:RK+RF}, Fig.~\ref{Fig:virusshare:CC+RandomForest} and Fig.~\ref{Fig:virusshare:CT+RandomForest}, model~\#7, \#8 and \#10, which are trained with the combinations of RAkELd and RandomForest, CC and RandomForest, CT and RandomForest, also have a similar tendency as model~\#1. For these 3 models, there is always such a situation that both the classification capability of the model and \# of samples for augmentation has no longer increased when the \# of batches reaches a certain number. For model~\#7, \#8, and \#10, the lift of the curves stops in epoch 5.
As shown in Fig.~\ref{Fig:virusshare:PS+RandomForest}, for model~\#6, after the first batch is added, the sample-ACC remains unchanged at 0.733. In epoch 2, \# of samples for augmentation starts to increase significantly. When the \# of batches equals 1,440 in epoch 3, 5,496 samples from the VirusShare dataset have been labeled with pseudo labels and added to the training set. Because the rest of the 4 samples other than the samples used for data augmentation cannot be successfully added, we consider the process of Detection-Training for this algorithm combination stops in 3 epochs.
In general, all these 5 processes of Detection-Training have proof of accuracy improvement and data augmentation, and can both be finished within 5 epochs, which means an acceptable efficiency.}

\paragraph{\textbf{Comparison between the DREBIN dataset and VirusShare dataset}}
Generally, the results on the two datasets reveal that the proposed active learning framework can not only improve the capability of MLC classifier but also construct a larger dataset with the help of the limited number of labeled malware samples. Besides the proof of accuracy improvement and data augmentation, we can observe that the processes of using the proposed Detection-Training on all five selected algorithm combinations can be finished within 6 epochs, which means an acceptable efficiency.

In detail, by comparing the results in Table.~\ref{tab:drebin:detection_training} and \ref{tab:virushare:detection_training}, 4 algorithm combinations, which are CDN and J48 (model~\#1), PS and RandomForest (model~\#6), CC and RandomForest (model~\#8), CT and RandomForest (model~\#10), perform good on both of the two selected datasets. From the aspect of improving accuracy, these 4 combinations achieve similar results with a difference of less than 0.034.
From the aspect of data augmentation, the combinations of CDN and J48 (model~\#1), PS, and RandomForest (model~\#6) achieve similar results on 2 datasets with differences at 544 and 40 samples, which only occupy 0.112 and 0.007 of the total on the lesser side (on DREBIN). However, the sizes of data augmentation achieved by the other two combinations on two datasets show a big difference at 1,592 and 632, which occupy 0.861 and 0.581 of the total on the lesser side (on the VirusShare dataset).
Besides, the combination of CDN and REPTree (model~\#4), RAkELd, and RandomForest (model~\#7) demonstrated significantly different effects on different datasets. These phenomena are quite interesting and worth future exploration. Through this work which is still in the early stages of this new direction, the biggest potentially possible reason is the variability of the datasets. As we introduced in previous sections (\S~\ref{intro} and \S~\ref{approach:detection_training}), it is quite hard and expensive to manually validate the malicious behaviors in malware, even with a detailed security report. In \S~\ref{discussion}, we will discuss some engineering approaches, which can potentially simplify the analysis process and are expected to be validated in future work. 

\begin{tcolorbox}[size=title,opacityfill=0.1,breakable]
\noindent\textit{\textbf{{Conclusion 2.1:}} {{By conducting the process of Detection-Training on {the DREBIN dataset} and VirusShare dataset, we confirm the effectiveness of our method. Specifically, for {the DREBIN dataset,} the best practice has enlarged the dataset to a size of {4,872}, and the sample-ACC reaches {0.867}. For the VirusShare dataset, \# of samples of the best practice is {5,416} and its sample-ACC is 0.833.}} }
\end{tcolorbox}

\subsubsection{Evaluation on Parameters Tuning in Detection-Training}
\label{evaluation:detection_training:parameter_tuning}
{To find out the best parameter setting of Detection-Training for better enhancing the effect of data augmentation and improving the classification capability on each selected algorithm combination, we conduct the experiments of tuning the batch size N on 2 datasets.}

\paragraph{\textbf{Dataset}}
\label{evaluation:detection_training:parameter_tuning:dataset}
{The dataset configuration used in this experiment is same as \S~\ref{evaluation:detection_training:active_learning:dataset_drebin} and \S~\ref{data:virusshare}}

\paragraph{\textbf{Experiment setup}}
\label{evaluation:detection_training:parameter_tuning:setup}
{As introduced in \S~\ref{evaluation:detection_training:active_learning:setup}, the default batch size and number of epochs are 8 and 6. In order to further explore the potential impact of different parameter settings, we pick 4, 8, 16, and 32 as the candidate batch sizes in this experiment, and determine the best number of epochs by plotting the result on accuracy improvement and data augmentation.
We pick 5 algorithm combinations for each unlabeled dataset (i.e., the DREIN dataset and VirusShare dataset), which has been initially proved to enlarge the training set with thousands of new malware samples as data augmentation, according to the results in \S~\ref{evaluation:detection_training:active_learning:drebin} and \S~\ref{evaluation:detection_training:active_learning:virusshare}.}

\paragraph{\textbf{Result on the DREBIN dataset}}
\label{evaluation:detection_training:parameter_tuning:drebin}

{We demonstrate and detail the results on our website~\cite{googleSites}. Overall, we can find that adopting 8 as batch size can achieve the highest classification accuracy and the largest size of data augmentation to the best practice for 3 combinations, such as CDN and J48, CDN and REPTree, PS, and RandomForest. For the other 2 combinations, the best batch size is 32. Besides, among all the models, whose data augmentation contains more than a thousand samples, model~\#1.2 has the best overall result in terms of all evaluation metrics.}

\paragraph{\textbf{Result on VirusShare dataset}}
\label{evaluation:detection_training:parameter_tuning:virusshare}

{We also demonstrate and detail the results on our website~\cite{googleSites}. In general, we can find that adopting 8 as batch size can achieve the highest classification accuracy and the largest size of data augmentation to the best practice for 4 combinations, such as CDN and J48, PS and RandomForest, CC and RandomForest, CT and RandomForest. This is the same as what we found in the experiment on the DREBIN dataset. For the combination of RAkELd and RandomForest, the best batch size is 32. Besides, among all the models, whose data augmentation contains more than a thousand samples, model~\#1.2 has the best result in terms of both \#
of samples and sample-ACC, proved to be an overall best model.}

\begin{tcolorbox}[size=title,opacityfill=0.1,breakable]
\noindent\textit{\textbf{{Conclusion 2.2:}} {By tuning the parameter (i.e., batch size) in Detection-Training, we can achieve the best practice of each algorithm combination on both data augmentation and classification capability. For most situations, when batch size is equal to 8, it can achieve excellent results within 6 epochs.}}
\end{tcolorbox}

\subsection{{RQ3: How reliable is the pseudo-label for the sample through Detection-Training?}}
\label{evaluation:label_valiadation}
{In this section, to verify the authenticity of the pseudo labels which are generated in the process of Detection-Training, we evaluate the effectiveness of MLC models which are trained with the pseudo-labeled samples on all sampled-based and label-based metrics.}

\subsubsection{Dataset}
\label{evaluation:label_valiadation:dataset}
{The dataset used in this experiment is same as \S~\ref{evaluation:detection_training:parameter_tuning:dataset}. However, we adopt an alternate dataset configuration in order to check the validity of the pseudo labels obtained within data augmentation. For each selected algorithm combination, we train the MLC classifier with a new training set, which is built from those pseudo-labeled malware samples obtained through Detection-Training (\S~\ref{evaluation:detection_training:active_learning}), and then test it with the manually labeled ground-truth dataset. Please do note that the pseudo-labeled malware samples used on each combination are strictly restricted to the generated data augmentation by the same algorithm combination.}

\subsubsection{Experiment setup}
\label{evaluation:label_valiadation:setup}
{The algorithm combinations used in this experiment are same as \S~\ref{evaluation:detection_training:parameter_tuning:setup}.}

\begin{table*}[ht]
\caption{Evaluation of models which use samples from the DREBIN dataset and VirusShare dataset}
\label{table:authenticity}
\centering
\setlength{\tabcolsep}{2.95mm}{
\begin{tabular}{|c|ccccc|ccccc|}
\hline
\begin{tabular}[c]{@{}c@{}}data\\ source\end{tabular}    & \multicolumn{5}{c|}{DREBIN}                                                                                                                                                                   & \multicolumn{5}{c|}{VirusShare}                                                                                                                                                               \\ \hline\hline
model id                                                   & 1     & 2       & 3                                                       & 4                                                       & 5                                                       & 1     & 2       & 3                                                       & 4                                                       & 5                                                       \\ \hline\hline
\begin{tabular}[c]{@{}c@{}}MLC\\ algorithm\end{tabular}    & CDN   & CDN     & PS                                                      & CC                                                      & CT                                                      & CDN   & PS     & RAkELd                                                      & CC                                                      & CT                                                      \\ \hline
\begin{tabular}[c]{@{}c@{}}basic\\ classifier\end{tabular} & J48   & \begin{tabular}[c]{@{}c@{}}{REPTree}\end{tabular} & \begin{tabular}[c]{@{}c@{}}Random\\ Forest\end{tabular} & \begin{tabular}[c]{@{}c@{}}Random\\ Forest\end{tabular} & \begin{tabular}[c]{@{}c@{}}Random\\ Forest\end{tabular} & J48   & \begin{tabular}[c]{@{}c@{}}Random\\ Forest\end{tabular} & \begin{tabular}[c]{@{}c@{}}Random\\ Forest\end{tabular} & \begin{tabular}[c]{@{}c@{}}Random\\ Forest\end{tabular} & \begin{tabular}[c]{@{}c@{}}Random\\ Forest\end{tabular} \\\hline\hline
hamming loss    & 0.169 & \cellcolor{graytbl}\textbf{0.091}    & 0.343 & {0.237} & {0.218} &  0.214 & \cellcolor{graytbl}\textbf{0.190}  &  0.248 &  0.245 &  0.229 \\
zero-one loss   & 0.646 &\cellcolor{graytbl}\textbf{0.393} & 0.882 & {0.888} & {0.787} & 0.843  &  \cellcolor{graytbl}\textbf{0.612} &   0.899&0.876   & 0.870  \\
F1-score    & 0.709 & \cellcolor{graytbl}\textbf{0.839}   & 0.354  &  {0.495}& {0.544} &  0.622 &\cellcolor{graytbl}\textbf{0.641} &  0.465 &  0.292 & 0.351  \\
sample-ACC  & 0.635 &  \cellcolor{graytbl}\textbf{0.781}   & 0.343  & {0.646} & {0.618} &   0.523 &   \cellcolor{graytbl}\textbf{0.697} &  0.669  &0.365  &  0.461  \\
Avg.label-ACC   & 0.831 &  \cellcolor{graytbl}\textbf{0.909} &  0.657   &   {0.763}  &   {0.782}&0.786  & \cellcolor{graytbl}\textbf{0.810} & 0.752  &0.755  & 0.772    \\
\hline
\end{tabular}
}
\end{table*}

\subsubsection{Results}
{In this section, we present the experiment results on the DREBIN dataset and VirusShare dataset referring to sample-based metrics and label-based metrics. We also discuss some insightful conclusions for relevant approaches using active learning and data augmentation.}

\paragraph{\textbf{Evaluation on the DREBIN dataset}}
\label{evaluation:label_valiadation:results:drebin}
{As shown on the left side of Table~\ref{table:authenticity}, model~\#2, which is trained with the combination of CDN and REPTree algorithms on the 5,456 pseudo-labeled samples from DREBIN, outperforms on all evaluation metrics. Compared with the result presents in Table.~\ref{tab:basemodel}, it performs better on all metrics than the base model which uses the same algorithm combination. Besides, compared with the updated model with the same combination (i.e., model~\#4 in Table.~\ref{tab:drebin:detection_training}) whose classification capability has been improved a lot through Detection-Training, it also achieves a better result in terms of all metrics. In detail, the hamming loss and zero-one loss have been reduced by 0.053 and 0.107, and the sample-ACC and averaged label-ACC have been improved by 0.014 and 0.053. So that we can make a conclusion that the 5,456 samples which are labeled using this combination during Detection-Training are well predicted. In other words, the generated pseudo labels are highly credible.
Model~\#1, which uses the combination of CDN and J48 that has the best result in \S~\ref{evaluation:detection_training:active_learning:drebin}, also performs well as the second best in this evaluation with the second lowest hamming loss and zero-one loss, and the second highest F1-score, sample-ACC and averaged label-ACC.
}

{We selected 20 samples and manually verify the authenticity of their pseudo-labels, the details are shown in Appendix B.}

\paragraph{\textbf{Evaluation on VirusShare dataset}}
{As shown on the right side of Table~\ref{table:authenticity}, model~\#2, which is trained with the combination of PS and RandomForest algorithms on the 5,496 pseudo-labeled samples from the VirusShare dataset, outperforms on all evaluation metrics. 
{Compared with the result presents in Table~\ref{tab:basemodel}, the sample-ACC is 0.003 less than its base model and the Avg.label-ACC is 0.057 less than that of its base model. 
Besides, compared with the updated model of the same combination (i.e., model~\#6 in Table.~\ref{tab:virushare:detection_training}), whose classification capability has been improved a lot through Detection-Training, it also has a lower classification accuracy. The reason for this phenomenon possibly is the samples from the VirusShare dataset collected in recent years (i.e. 2018-2022) may include some variants which use new methods to perform malicious behaviors.}
Model~\#3, which uses the combination of RAkELd and RandomForest that has the second best result in \S~\ref{evaluation:detection_training:active_learning:virusshare}, also performs well as the second best in this evaluation with a sample-ACC at 0.669.
Model~\#1, which uses the combination of CDN and J48 that has the best result in \S~\ref{evaluation:detection_training:active_learning:virusshare}, performs as the third best in this evaluation with the second lowest hamming loss and zero-one loss, the second highest F1-score and averaged label-ACC, and the third highest sample-ACC.
}
{{We also select 20 samples and manually verify the authenticity of their pseudo-labels, the details are shown in Appendix B.}}

\paragraph{\textbf{Comparison between DREBIN and VirusShare dataset}}
By comparing the results shown in Table~\ref{table:authenticity} with the results of the base model in Table~\ref{tab:basemodel}, we observe that the overall effectiveness of the proposed approach is validated to be successful because we achieve a significant improvement in all evaluation metrics with the model of CDN and RandomForest trained on pseudo-labeled samples from DREBIN. The result of best practice outperforms its based model by adopting pseudo-labeled samples as a training set and the manually labeled dataset as a test set.

Besides, we also notice some interesting results. For instance, the results on DREBIN are better than that on the VirusShare dataset for most same algorithm combinations, except the combination of PS and RandomForest. And, for the combinations of CDN and RandomForest, RAkELd, and RandomForest, the results of Detection-Training are diametrically opposed between the two datasets. The most possible reason that causes this phenomenon could be the distribution of the dataset on our defined malicious behaviors since DREBIN is a more established family dataset with higher diversity. Another one is about the adaptability of MLC and BC algorithms to our task.
Another example is that the sample-ACC of model~\#2 is a little lower than the base model which uses the same algorithm combination. One of the possible reasons behind this phenomenon could be the samples from the VirusShare dataset collected in recent years (i.e. 2018-2022) may include some variants which use new methods to perform malicious behaviors. Hence, the features of these behaviors may be not included in our feature set, leading to a miss classification.
In \S~\ref{discussion}, we will discuss some potential directions that can further explore the reasons behind these phenomena and the underlying principles.

\begin{tcolorbox}[size=title,opacityfill=0.1,breakable]
\noindent\textit{\textbf{{Conclusion:}} {{By evaluating the effectiveness of MLC models which are trained with pseudo-labeled samples, we can tell the confidence of generated pseudo labels. The best practice (0.781), which is trained on pseudo-labeled samples from {the DREBIN dataset}, can even outperform the best-based models (0.733), which is trained on the ground-truth dataset.}
}}
\end{tcolorbox}

\section{Related Work}
\subsection{Android Malware Classification}
{Facing the increasing number of malicious apps, malware classification is becoming extremely important. Malware Classification can refer either to the classification of binaries as malicious or benign, or the classification of malware samples into different known malware families. Towards android malware binary classification, many machine learning methods have achieved great success in Android malware~\cite{aafer2013droidapiminer,chen2018automated,chen2016stormdroid,chen2016towards,fan2016poster,fereidooni2016anastasia}. Many machine learning algorithms such as SVM~\cite{arp2014drebin} and Random forest~\cite{rastogi2013droidchameleon} were used to detect malware and have proven to be effective. The method presented by Yerima et al.~\cite{yerima2013new} detects Android malware based on Bayesian Classification models obtained from API calls, system commands, and permissions. Wu et al.~\cite{wu2016effective} used data-flow APIs as classification features to detect Android malware. With the popularity of deep neural networks, researchers utilized the deep neural network models for malware detection~\cite{feng_icfem,feng_mobidroid_2019,feng_seqmobile_2020,feng_performance_sensitive_2021,kim2018multimodal}.}

{For malware family classification, which classifies malware with common features into malware families, has been proposed as an effective malware analysis method. H. Peng et al.~\cite{peng2012using} proposed a family classification method using permissions as features, they introduced a probabilistic generative model to rank the risks of Android applications. M. Zhang et al.~\cite{zhang2014semantics} proposed a novel semantic-based approach that classifies Android malware via dependency graphs. M. Fan et al.~\cite{fan2018android} proposed an approach that constructs frequent subgraphs to represent the common behaviors of malware samples that belong to the same family for enhancing the correctness of family classification. Y. Bai et al.~\cite{bai2020icse} proposed a novel siamese-network based learning method for malware family classification.}

{Unlike the existing approaches on binary and family classification above, we highlight that this is the first multi-label Android malware classification approach for each malware which intends to provide more information on fine-grained malicious behaviors through detection results.}

\subsection{Multi-label Classification} 
{Generally, multi-label classification (MLC) aims to attach multiple relevant labels to an input instance simultaneously through machine learning. During the past decades, multi-label classification has been widely adopted by communities in various domains~\cite{zhouzhihua_MLC_2014}, such as natural language processing, web mining, computer vision, bioinformatics, and multimedia. MLC has mainly engaged the attention of researchers working on text categorization~\cite{NIPS2002_3147da8a,NIPS2004_0bed45bd}, as each member of a document collection usually belongs to more than one semantic category. In web mining, various problems brought by the massive information are also discovered with MLC, such as page categorization and tag suggestion~\cite{katakis2008multilabel,tag_suggestion_2008,Metalabeler}. In image semantic classification, a common scenario may contain multiple objects. With MLC, researchers can annotate the given scene with multiple labels, which makes scene classification more comprehensive~\cite{boutell2004learning,kang2006correlated}. MLC is also applied to the field of protein function classification, as there are proteins that have multiple properties at the same time or exist in multiple subcellular organelles. Researchers use multiple labels on those data for proteins and achieve promising results with MLC~\cite{clare2001knowledge}. In the field of multimedia, for the tasks that retrieve relevant (categories) emotions associated with a given piece of music, MLC can achieve success by learning about the music with multiple emotional labels~\cite{li2003detecting,trohidis2008multi}. In our task, there exists the same situation as the above problems from other fields, which is one malware may have multiple malicious behaviors and some of them beyond their malware families' definitions. The original idea of our work is inspired by the above MLC-based work from other domains.}

\subsection{Data Augmentation and Active Learning} 
%active learning
{Most supervised learning-based methods tend to require immense amounts of computational and human resources for data annotation so designing effective methods that can enlarge a small labeled dataset without too much effort on annotations is a fundamental research challenge in many classification tasks based on big data. In the past decade, many presented research work using semi-supervised methods mainly focusing on algorithm innovation, engineering practice, and new application exploration. Along with this style of research and development, recently, data augmentation and active learning have been proved that it can successfully alleviate the data-deficient scenarios in many research fields, such as natural language processing~\cite{al_nlp} and medical image analysis~\cite{al_medical}.
For data augmentation, SPADE~\cite{phamspade} used a two-phase data augmentation process to enrich a dataset before training a deep-learning classifier. D.H. Lee et al.~\cite{lee2013pseudo} picked the labels with the largest predicted probability to generate pseudo labels for unlabeled data. Mixmatch~\cite{berthelot2019mixmatch} applied the mix-up technique to both pseudo-labeled data and unlabeled samples, then use these data to iteratively train the network. L. Samuli~\cite{laine2016temporal} formed a consensus prediction of the unknown labels using the outputs of the network-in-training on different epochs. T. Antti~\cite{tarvainen2017mean} took a weight-average of the parameters of the model and used this weight-averaged parameter to generate pseudo labels. 
For active learning, J. Zhang et al.~\cite{zhang2021self} applied active learning and data augmentation to the field of anomaly detection, the NAF-AL method they proposed uses the augmented samples incorporated with normal samples to train a better anomaly detector. Z. Zhou et al.~\cite{zhou2021active} proposed an active learning approach based on data augmentation for reducing annotation efforts in medical image analysis. In the ED2~\cite{al_error}, active learning was applied in the error detection field, which aims to solve a fundamental problem in data science. }

{Data augmentation and Active learning also have several applications in the field of malware detection. 
For data augmentation, Ding et al.~\cite{data-augment-yuxu} proposed a Grad-CAM algorithm to find the raw data representing malware features and generated malware family samples. Chen et al.~\cite{data-augment-chen} first transformed malware into images and then used the generative adversarial networks to generate data. The generated data can be used to balance and expand the original dataset. For Active learning, Chen et al.~\cite{active-learning} leveraged the Active learning by learning (ALBL) technique based on the experts' feedback. So that the trained models can be updated to reduce the classification error rate.}

As we introduced in \S~\ref{intro} and \S~\ref{approach:detection_training}, we also face a similar problem in that the number of usable labeled data is limited and the cost of effort in data annotation is unbearable. Inspired by the above work from other domains, we take the idea of active learning and data augmentation to generate pseudo labels for unlabeled samples, iteratively retrain models to validate the correctness of pseudo labels as well as improve the capability of classification.

\section{Discussion and Future Work}\label{discussion}
As we discussed in the introduction (\S~\ref{intro}), there exist (1) malicious behaviors that may be achieved with various key features; (2) malware that has more diverse kinds of malicious behaviors beyond the 6 label; (3) the best classification effectiveness cannot be achieved with API, permission, intent against the implementation of some behaviors; (4) changing of some APIs within the update of Android SDK. All these reasons lead to the incomplete of our defined label set (as well as the used features), which is the first threat in our work. To tackle this threat, we will (1) supplement the malicious behavior label categories and organize a summary of more potential ways to achieve identical malicious behaviors by analyzing more malware collected in recent years; (2) try other feature representation methods, such as call graph, etc, for more context information, which is beneficial to overcome the current limitation of ``feature overlap''; (3) systematically update APIs in the feature dictionary to meet different Android system versions.
As the experiment of the authenticity validation on pseudo labels (\S~\ref{evaluation:label_valiadation}) shows, the pseudo labels of involved samples from the VirusShare dataset show less credibility compared to those from the DREBIN dataset. That is possible because (1) the VirusShare data are collected in recent years, which may include variants which use new approaches to perform malicious behaviors; (2) the dataset may be imbalanced on different labels; (3) training the base model with limited data can make the collected pseudo-labeled samples imbalanced. This is the second threat in our consideration. To face this, we will look into the technologies from other domains for (1) saving efforts on manual analysis and data annotation for constructing a relatively balanced and original dataset; (2) automatically verifying the labels in the process of Detection-Training with open-access security reports (e.g., confidence) and the changes in numerical information (e.g., variance, entropy) during active learning.
Besides the threats, we will also try to further explore the potential application of multi-label classification in the field of malware detection, such as AI malicious behaviors interpretation based on MLC.

\section{Conclusion}
{This paper is an initial attempt to study the multi-label classification problem of malicious behaviors in Android malware classification. We come up with a label set of 6 types of malicious behaviors, and manually label 180 malware with an in-depth study, which is released as a benchmark. To address the challenge in data annotation, we demonstrate a novel active learning method based on data augmentation and achieve a promising result.
}

%\section*{Acknowledgment}
%This work was partially supported by the National Natural Science Foundation of China under Grant No. 62102284, No. 61872262, Smart Platform Infrastructure Research on Integrative Technology (SPIRIT) under Grant CSA-CSEC-DC-20-083, and Software Security with a Focus on Critical Technologies (Socrates) under Grant C21-00267 sponsored by Cyber Security CRC.

\bibliographystyle{IEEEtran}
\bibliography{reference}

% bio
%\input{bio}

% \break
% \input{appendix.tex}

\end{document}